\documentclass[longauth]{aa}
\usepackage{graphicx}
\usepackage{subfigure}
\usepackage{txfonts}
\usepackage{siunitx}
\usepackage{xspace}
\usepackage[switch]{lineno}
\usepackage[colorlinks,allcolors=blue]{hyperref}
\bibpunct{(}{)}{;}{a}{}{,}

\newcommand{\gam}{$\gamma$\xspace}
\newcommand{\gp}{\textsc{Gammapy}\xspace}
\newcommand{\fermipy}{\textsc{Fermipy}\xspace}
\newcommand{\hess}{H.E.S.S.\xspace}
\newcommand{\fermi}{\textit{Fermi}-LAT\xspace}
\newcommand{\Fermi}{\textit{Fermi}~LAT\xspace}
\newcommand{\vbm}{variable $B$-field model\xspace}
\newcommand{\cbm}{constant $B$-field model\xspace}
\newcommand{\RNum}[1]{\uppercase\expandafter{\romannumeral #1\relax}}
\graphicspath{{figures/}{./}}

\makeatletter
\renewcommand*\aa@pageof{, page \thepage{} of \pageref*{LastPage}}
\makeatother

\DeclareSIUnit\gauss{G}

\begin{document}

\title{Spectrum and extension of the inverse-Compton emission of the Crab Nebula from a combined \fermi and \hess analysis}
\titlerunning{Spectrum and extension of the Crab Nebula from a combined \fermi and \hess analysis}

\author{\begin{small}
F.~Aharonian \inst{\ref{DIAS},\ref{MPIK},\ref{Yerevan}}
\and F.~Ait~Benkhali \inst{\ref{LSW}}
\and J.~Aschersleben \inst{\ref{Groningen}}
\and H.~Ashkar \inst{\ref{LLR}}
\and M.~Backes \inst{\ref{UNAM},\ref{NWU}}
\and A.~Baktash \inst{\ref{UHH}}
\and V.~Barbosa~Martins \inst{\ref{DESY}}
\and R.~Batzofin \inst{\ref{UP}}
\and Y.~Becherini \inst{\ref{APC},\ref{Linnaeus}}
\and D.~Berge \inst{\ref{DESY},\ref{HUB}}
\and K.~Bernl\"ohr \inst{\ref{MPIK}}
\and B.~Bi \inst{\ref{IAAT}}
\and M.~B\"ottcher \inst{\ref{NWU}}
\and C.~Boisson \inst{\ref{LUTH}}
\and J.~Bolmont \inst{\ref{LPNHE}}
\and M.~de~Bony~de~Lavergne \inst{\ref{CEA}}
\and J.~Borowska \inst{\ref{HUB}}
\and F.~Bradascio \inst{\ref{CEA}}
\and M.~Breuhaus \inst{\ref{MPIK}}
\and R.~Brose \inst{\ref{DIAS}}
\and A.~Brown \inst{\ref{Oxford}}
\and F.~Brun \inst{\ref{CEA}}
\and B.~Bruno \inst{\ref{ECAP}}
\and T.~Bulik \inst{\ref{UWarsaw}}
\and C.~Burger-Scheidlin \inst{\ref{DIAS}}
\and T.~Bylund \inst{\ref{CEA}}
\and S.~Caroff \inst{\ref{LAPP}}
\and S.~Casanova \inst{\ref{IFJPAN}}
\and R.~Cecil \inst{\ref{UHH}}
\and J.~Celic \inst{\ref{ECAP}}
\and M.~Cerruti \inst{\ref{APC}}
\and P.~Chambery \inst{\ref{CENBG}}
\and T.~Chand \inst{\ref{NWU}}
\and S.~Chandra \inst{\ref{NWU}}
\and A.~Chen \inst{\ref{Wits}}
\and J.~Chibueze \inst{\ref{NWU}}
\and O.~Chibueze \inst{\ref{NWU}}
\and G.~Cotter \inst{\ref{Oxford}}
\and P.~Cristofari \inst{\ref{LUTH}}
\and J.~Devin \inst{\ref{LUPM}}
\and A.~Djannati-Ata\"i \inst{\ref{APC}}
\and J.~Djuvsland \inst{\ref{MPIK}}
\and A.~Dmytriiev \inst{\ref{NWU}}
\and S.~Einecke \inst{\ref{Adelaide}}
\and J.-P.~Ernenwein \inst{\ref{CPPM}}
\and S.~Fegan \inst{\ref{LLR}}
\and K.~Feijen \inst{\ref{APC}}
\and M.~Filipovi\'c \inst{\ref{WSU}}
\and G.~Fontaine \inst{\ref{LLR}}
\and M.~F\"u\ss{}ling \inst{\ref{DESY}}
\and S.~Funk \inst{\ref{ECAP}}
\and S.~Gabici \inst{\ref{APC}}
\and Y.A.~Gallant \inst{\ref{LUPM}}
\and G.~Giavitto \inst{\ref{DESY}}
\and D.~Glawion \inst{\ref{ECAP}}
\and J.F.~Glicenstein \inst{\ref{CEA}}
\and J.~Glombitza \inst{\ref{ECAP}}
\and P.~Goswami \inst{\ref{APC}}
\and G.~Grolleron \inst{\ref{LPNHE}}
\and M.-H.~Grondin \inst{\ref{CENBG}}
\and L.~Haerer \inst{\ref{MPIK}}
\and J.A.~Hinton \inst{\ref{MPIK}}
\and W.~Hofmann \inst{\ref{MPIK}}
\and T.~L.~Holch \inst{\ref{DESY}}
\and M.~Holler \inst{\ref{Innsbruck}}
\and D.~Horns \inst{\ref{UHH}}
\and M.~Jamrozy \inst{\ref{UJK}}
\and F.~Jankowsky \inst{\ref{LSW}}
\and V.~Joshi \inst{\ref{ECAP}}
\and E.~Kasai \inst{\ref{UNAM}}
\and K.~Katarzy\'nski \inst{\ref{NCUT}}
\and R.~Khatoon \inst{\ref{NWU}}
\and B.~Kh\'elifi \inst{\ref{APC}}
\and W.~Klu\'zniak \inst{\ref{NCAC}}
\and Nu.~Komin \inst{\ref{Wits}}
\and K.~Kosack \inst{\ref{CEA}}
\and D.~Kostunin \inst{\ref{DESY}}
\and A.~Kundu \inst{\ref{NWU}}
\and R.G.~Lang \inst{\ref{ECAP}}
\and S.~Le~Stum \inst{\ref{CPPM}}
\and F.~Leitl \inst{\ref{ECAP}}
\and A.~Lemi\`ere \inst{\ref{APC}}
\and M.~Lemoine-Goumard \inst{\ref{CENBG}}
\and J.-P.~Lenain \inst{\ref{LPNHE}}
\and F.~Leuschner \inst{\ref{IAAT}}
\and A.~Luashvili \inst{\ref{LUTH}}
\and J.~Mackey \inst{\ref{DIAS}}
\and D.~Malyshev \inst{\ref{IAAT}}
\and D.~Malyshev \inst{\ref{ECAP}}
\and V.~Marandon \inst{\ref{CEA}}
\and P.~Marinos \inst{\ref{Adelaide}}
\and G.~Mart\'i-Devesa \inst{\ref{Innsbruck}}
\and R.~Marx \inst{\ref{LSW}}
\and A.~Mehta \inst{\ref{DESY}}
\and M.~Meyer \inst{\ref{UHH}}$^{,}$\footnotemark[1] 
\and A.~Mitchell \inst{\ref{ECAP}}
\and R.~Moderski \inst{\ref{NCAC}}
\and L.~Mohrmann \inst{\ref{MPIK}}$^{,}$\thanks{Corresponding authors;\\\email{\href{mailto:contact.hess@hess-experiment.eu}{contact.hess@hess-experiment.eu}}}
\and A.~Montanari \inst{\ref{LSW}}
\and E.~Moulin \inst{\ref{CEA}}
\and T.~Murach \inst{\ref{DESY}}
\and M.~de~Naurois \inst{\ref{LLR}}
\and J.~Niemiec \inst{\ref{IFJPAN}}
\and P.~O'Brien \inst{\ref{Leicester}}
\and S.~Ohm \inst{\ref{DESY}}
\and L.~Olivera-Nieto \inst{\ref{MPIK}}
\and E.~de~Ona~Wilhelmi \inst{\ref{DESY}}
\and M.~Ostrowski \inst{\ref{UJK}}
\and S.~Panny \inst{\ref{Innsbruck}}
\and M.~Panter \inst{\ref{MPIK}}
\and R.D.~Parsons \inst{\ref{HUB}}
\and G.~Peron \inst{\ref{APC}}
\and D.A.~Prokhorov \inst{\ref{Amsterdam}}
\and G.~P\"uhlhofer \inst{\ref{IAAT}}
\and M.~Punch \inst{\ref{APC}}
\and A.~Quirrenbach \inst{\ref{LSW}}
\and M.~Regeard \inst{\ref{APC}}
\and P.~Reichherzer \inst{\ref{CEA}}
\and A.~Reimer \inst{\ref{Innsbruck}}
\and O.~Reimer \inst{\ref{Innsbruck}}
\and H.~Ren \inst{\ref{MPIK}}
\and M.~Renaud \inst{\ref{LUPM}}
\and B.~Reville \inst{\ref{MPIK}}
\and F.~Rieger \inst{\ref{MPIK}}
\and G.~Roellinghoff \inst{\ref{ECAP}}
\and B.~Rudak \inst{\ref{NCAC}}
\and V.~Sahakian \inst{\ref{Yerevan_Phys}}
\and H.~Salzmann \inst{\ref{IAAT}}
\and M.~Sasaki \inst{\ref{ECAP}}
\and F.~Sch\"ussler \inst{\ref{CEA}}
\and H.M.~Schutte \inst{\ref{NWU}}
\and J.N.S.~Shapopi \inst{\ref{UNAM}}
\and A.~Specovius \inst{\ref{ECAP}}
\and S.~Spencer \inst{\ref{ECAP}}
\and {\L.}~Stawarz \inst{\ref{UJK}}
\and R.~Steenkamp \inst{\ref{UNAM}}
\and S.~Steinmassl \inst{\ref{MPIK}}
\and C.~Steppa \inst{\ref{UP}}
\and K.~Streil \inst{\ref{ECAP}}
\and I.~Sushch \inst{\ref{NWU}}
\and H.~Suzuki \inst{\ref{Konan}}
\and T.~Takahashi \inst{\ref{KAVLI}}
\and T.~Tanaka \inst{\ref{Konan}}
\and R.~Terrier \inst{\ref{APC}}
\and M.~Tluczykont \inst{\ref{UHH}}
\and N.~Tsuji \inst{\ref{RIKKEN}}
\and T.~Unbehaun \inst{\ref{ECAP}}$^{,}$\footnotemark[1] 
\and C.~van~Eldik \inst{\ref{ECAP}}
\and M.~Vecchi \inst{\ref{Groningen}}
\and J.~Veh \inst{\ref{ECAP}}
\and C.~Venter \inst{\ref{NWU}}
\and J.~Vink \inst{\ref{Amsterdam}}
\and T.~Wach \inst{\ref{ECAP}}
\and S.J.~Wagner \inst{\ref{LSW}}
\and A.~Wierzcholska \inst{\ref{IFJPAN}}
\and M.~Zacharias \inst{\ref{LSW},\ref{NWU}}
\and D.~Zargaryan \inst{\ref{DIAS}}
\and A.A.~Zdziarski \inst{\ref{NCAC}}
\and A.~Zech \inst{\ref{LUTH}}
\and S.~Zouari \inst{\ref{APC}}
\and N.~\.Zywucka \inst{\ref{NWU}} (\hess Collaboration)
\and A.~Harding \inst{\ref{LosAlamos}} \end{small}
}
\authorrunning{F. Aharonian et al.}

\institute{
Dublin Institute for Advanced Studies, 31 Fitzwilliam Place, Dublin 2, Ireland \label{DIAS} \and
Max-Planck-Institut f\"ur Kernphysik, P.O. Box 103980, D 69029 Heidelberg, Germany \label{MPIK} \and
Yerevan State University,  1 Alek Manukyan St, Yerevan 0025, Armenia \label{Yerevan} \and
Landessternwarte, Universit\"at Heidelberg, K\"onigstuhl, D 69117 Heidelberg, Germany \label{LSW} \and
Kapteyn Astronomical Institute, University of Groningen, Landleven 12, 9747 AD Groningen, The Netherlands \label{Groningen} \and
Laboratoire Leprince-Ringuet, École Polytechnique, CNRS, Institut Polytechnique de Paris, F-91128 Palaiseau, France \label{LLR} \and
University of Namibia, Department of Physics, Private Bag 13301, Windhoek 10005, Namibia \label{UNAM} \and
Centre for Space Research, North-West University, Potchefstroom 2520, South Africa \label{NWU} \and
Universit\"at Hamburg, Institut f\"ur Experimentalphysik, Luruper Chaussee 149, D 22761 Hamburg, Germany \label{UHH} \and
Deutsches Elektronen-Synchrotron DESY, Platanenallee 6, 15738 Zeuthen, Germany \label{DESY} \and
Institut f\"ur Physik und Astronomie, Universit\"at Potsdam,  Karl-Liebknecht-Strasse 24/25, D 14476 Potsdam, Germany \label{UP} \and
Université de Paris, CNRS, Astroparticule et Cosmologie, F-75013 Paris, France \label{APC} \and
Department of Physics and Electrical Engineering, Linnaeus University,  351 95 V\"axj\"o, Sweden \label{Linnaeus} \and
Institut f\"ur Physik, Humboldt-Universit\"at zu Berlin, Newtonstr. 15, D 12489 Berlin, Germany \label{HUB} \and
Institut f\"ur Astronomie und Astrophysik, Universit\"at T\"ubingen, Sand 1, D 72076 T\"ubingen, Germany \label{IAAT} \and
Laboratoire Univers et Théories, Observatoire de Paris, Université PSL, CNRS, Université Paris Cité, 5 Pl. Jules Janssen, 92190 Meudon, France \label{LUTH} \and
Sorbonne Universit\'e, Universit\'e Paris Diderot, Sorbonne Paris Cit\'e, CNRS/IN2P3, Laboratoire de Physique Nucl\'eaire et de Hautes Energies, LPNHE, 4 Place Jussieu, F-75252 Paris, France \label{LPNHE} \and
IRFU, CEA, Universit\'e Paris-Saclay, F-91191 Gif-sur-Yvette, France \label{CEA} \and
University of Oxford, Department of Physics, Denys Wilkinson Building, Keble Road, Oxford OX1 3RH, UK \label{Oxford} \and
Friedrich-Alexander-Universit\"at Erlangen-N\"urnberg, Erlangen Centre for Astroparticle Physics, Nikolaus-Fiebiger-Str. 2, 91058 Erlangen, Germany \label{ECAP} \and
Astronomical Observatory, The University of Warsaw, Al. Ujazdowskie 4, 00-478 Warsaw, Poland \label{UWarsaw} \and
Université Savoie Mont Blanc, CNRS, Laboratoire d'Annecy de Physique des Particules - IN2P3, 74000 Annecy, France \label{LAPP} \and
Instytut Fizyki J\c{a}drowej PAN, ul. Radzikowskiego 152, 31-342 Krak{\'o}w, Poland \label{IFJPAN} \and
Universit\'e Bordeaux, CNRS, LP2I Bordeaux, UMR 5797, F-33170 Gradignan, France \label{CENBG} \and
School of Physics, University of the Witwatersrand, 1 Jan Smuts Avenue, Braamfontein, Johannesburg, 2050 South Africa \label{Wits} \and
Laboratoire Univers et Particules de Montpellier, Universit\'e Montpellier, CNRS/IN2P3,  CC 72, Place Eug\`ene Bataillon, F-34095 Montpellier Cedex 5, France \label{LUPM} \and
School of Physical Sciences, University of Adelaide, Adelaide 5005, Australia \label{Adelaide} \and
Aix Marseille Universit\'e, CNRS/IN2P3, CPPM, Marseille, France \label{CPPM} \and
School of Science, Western Sydney University, Locked Bag 1797, Penrith South DC, NSW 2751, Australia \label{WSU} \and
Universit\"at Innsbruck, Institut f\"ur Astro- und Teilchenphysik, Technikerstraße 25, 6020 Innsbruck, Austria \label{Innsbruck} \and
Obserwatorium Astronomiczne, Uniwersytet Jagiello{\'n}ski, ul. Orla 171, 30-244 Krak{\'o}w, Poland \label{UJK} \and
Institute of Astronomy, Faculty of Physics, Astronomy and Informatics, Nicolaus Copernicus University,  Grudziadzka 5, 87-100 Torun, Poland \label{NCUT} \and
Nicolaus Copernicus Astronomical Center, Polish Academy of Sciences, ul. Bartycka 18, 00-716 Warsaw, Poland \label{NCAC} \and
Department of Physics and Astronomy, The University of Leicester, University Road, Leicester, LE1 7RH, United Kingdom \label{Leicester} \and
GRAPPA, Anton Pannekoek Institute for Astronomy, University of Amsterdam,  Science Park 904, 1098 XH Amsterdam, The Netherlands \label{Amsterdam} \and
Yerevan Physics Institute, 2 Alikhanian Brothers St., 0036 Yerevan, Armenia \label{Yerevan_Phys} \and
Department of Physics, Konan University, 8-9-1 Okamoto, Higashinada, Kobe, Hyogo 658-8501, Japan \label{Konan} \and
Kavli Institute for the Physics and Mathematics of the Universe (WPI), The University of Tokyo Institutes for Advanced Study (UTIAS), The University of Tokyo, 5-1-5 Kashiwa-no-Ha, Kashiwa, Chiba, 277-8583, Japan \label{KAVLI} \and
RIKEN, 2-1 Hirosawa, Wako, Saitama 351-0198, Japan \label{RIKKEN} \and
Theoretical Division, Los Alamos National Laboratory Los Alamos, NM 87545 USA \label{LosAlamos}
}

\abstract{
    The Crab Nebula is a unique laboratory for studying the acceleration of electrons and positrons through their non-thermal radiation.
    Observations of very-high-energy \gam rays from the Crab Nebula have provided important constraints for modelling its broadband emission.
    We present the first fully self-consistent analysis of the Crab Nebula's \gam-ray emission between \SI{1}{\GeV} and $\sim$\SI{100}{\TeV}, that is, over five orders of magnitude in energy.
    Using the open-source software package \gp, we combined \SI{11.4}{yr} of data from the \emph{Fermi} Large Area Telescope and \SI{80}{h} of High Energy Stereoscopic System (\hess) data at the event level and provide a measurement of the spatial extension of the nebula and its energy spectrum.
    We find evidence for a shrinking of the nebula with increasing \gam-ray energy.
    Furthermore, we fitted several phenomenological models to the measured data, finding that none of them can fully describe the spatial extension and the spectral energy distribution at the same time.
    Especially the extension measured at TeV energies appears too large when compared to the X-ray emission.
    Our measurements probe the structure of the magnetic field between the pulsar wind termination shock and the dust torus, and we conclude that the magnetic field strength decreases with increasing distance from the pulsar.
    We complement our study with a careful assessment of systematic uncertainties.
}


\maketitle

\section{Introduction}
The Crab Nebula, associated with the pulsar PSR~B0531+21, represents the archetype of a pulsar wind nebula and has been extensively studied across all wavelengths of the electromagnetic spectrum \citep[e.g.][]{Hester2008,Buehler2014,Amato2021}.
In the high-energy (HE; $\SI{100}{MeV}<E<\SI{100}{GeV}$) and very-high-energy (VHE; $E>\SI{100}{GeV}$) wavebands, it stands out as one of the brightest objects in the sky.
In fact, the Crab Nebula was the first \gam-ray source to be unambiguously detected at very high energies from 1986--1988 with the Whipple telescope \citep{Weekes1989}.
The radiation emitted by the nebula is mostly steady, although temporal variations have been observed in some wavebands \citep[e.g.][]{WilsonHodge2011,FermiLAT_CrabFlares_2011}.
In the VHE band, no flux variability has been observed so far, and the Crab Nebula has been used as a standard candle for detector calibration in this regime since its first detection \citep[cf.][]{Vacanti1991,HEGRA2003,HESS2006,MAGIC2008,Meyer2010,HAWC2017,LHAASO2021_CrabPerformance,CTA_LST_2023}.

\begin{figure*}
    \includegraphics[width=0.99\textwidth]{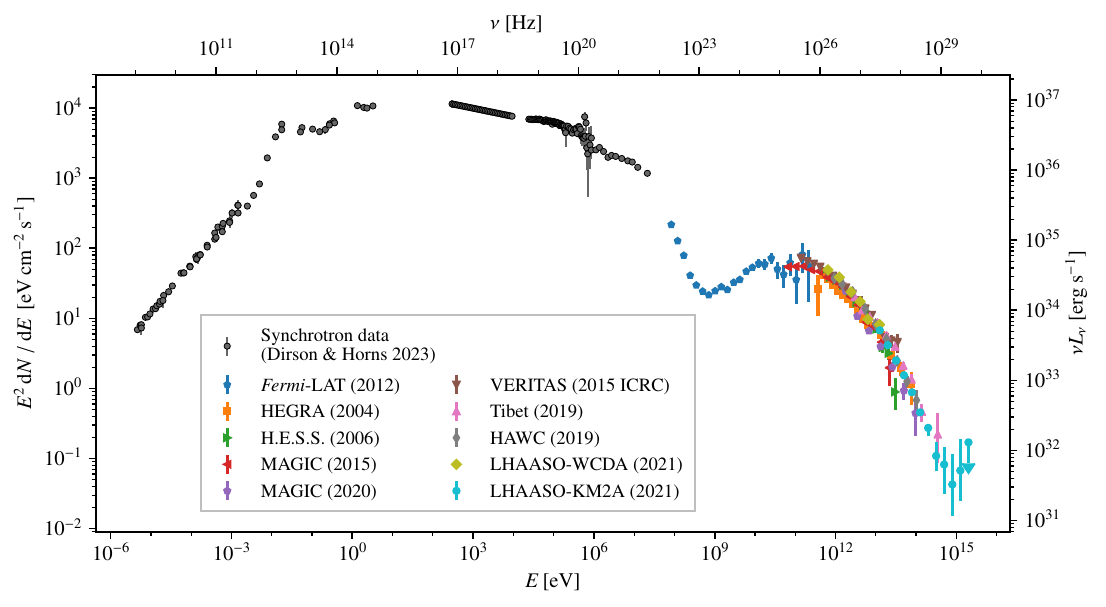}
    \caption{
        SED of the Crab Nebula.
        The synchrotron data points up to hard X-rays have been taken from \citet{Dirson2023} and references therein.
        The HE and VHE data are from \citet{Buehler2012}, \citet{HEGRA2004}, \citet{HESS2006}, \citet{MAGIC2015}, \citet{MAGIC2020}, \citet{VERITAS2015}, \citet{Tibet2019}, \citet{HAWC2019}, and \citet{LHAASO2021}, in the order as listed in the legend.
        A distance of \SI{2}{kpc} has been assumed to convert to luminosity (see right-hand side axis).
    }
    \label{fig:sed_intro}
\end{figure*}

We show in Fig.~\ref{fig:sed_intro} the spectral energy distribution (SED) of the Crab Nebula, as measured by a large number of instruments.
The distribution exhibits two broad peaks; both are generally accepted to be caused by a population of high-energy electrons\footnote{Throughout this paper, the term `electrons' refers to both electrons and positrons.} that fills the nebula.
The first peak is due to synchrotron emission that the electrons emit within the magnetic field of the nebula.
The emission at energies close to the second peak is generated by high-energy electrons via the inverse Compton (IC) process.
The general formation and structure of the nebula was already discussed by \citet{Rees1974}, and a number of theoretical models describing the emission from high-energy electrons in the Crab Nebula have been put forward since then \citep[e.g.][]{Kennel1984a,deJager1992,Atoyan1996,Hillas1998,HEGRA2004,Meyer2010}, including time-dependent models applicable to young pulsar wind nebulae (e.g.\ \citealt{Torres2014,vanRensburg2018}; for a review see also \citealt{Amato2021} and references therein).
Although usually considered sub-dominant over most of the energy range, \gam-ray emission due to accelerated cosmic-ray nuclei has also been considered \citep[e.g.][]{Atoyan1996,Arons1998,Bednarek2003}.
Recently, following the detection of PeV-energy photons from the Crab Nebula by the LHAASO detector \citep{LHAASO2021}, this possibility has received some renewed interest \citep{Peng2022,Nie2022}.

In recent years, it has furthermore become possible to spatially resolve the IC component of the Crab Nebula both in the HE and VHE bands, with extensions (68\% containment radii) of $1.8'\pm 0.18_\mathrm{stat}'\pm 0.42_\mathrm{sys}'$ measured with the \emph{Fermi} Large Area Telescope \citep[LAT;][]{FermiLAT_ExtendedSources_2018} and $1.31'\pm 0.07_\mathrm{stat}' \pm 0.17_\mathrm{sys}'$ measured with the High Energy Stereoscopic System \citep[H.E.S.S.;][]{HESS2020}, respectively.
These measurements provide new, crucial constraints for the theoretical modelling of the IC emission from the Crab Nebula.
Generally, because higher-energy electrons cool more efficiently, the extension of the nebula is expected to decrease with increasing electron energy \citep{Kennel1984a,deJager1992,Atoyan1996,Hillas1998,HEGRA2004}.
For example, in the X-ray domain -- where the emission is due to synchrotron radiation of the highest-energy electrons in the system -- the nebula exhibits a torus-like morphology that is only about $0.63'\times 0.3'$ in size \citep{Weisskopf2000}.
In contrast, in the ultraviolet, the characteristic extension is approximately $1.6'$ \citep[][where we have converted from a containment fraction of 39\% to 68\% assuming a Gaussian morphology]{HESS2020}.
However, besides the spatial distribution of the electrons, the apparent size of the nebula also depends on the extension of the nebula's magnetic field (for synchrotron radiation) and its distribution of seed photons (for the IC component).
At radio wavelengths the extension is indeed only $\sim$$1.9'$, which is smaller than expected solely from the distribution of electrons \citep{Yeung2019}.
Recently, \citet{Dirson2023} performed a combined spatial and spectral fit of the multi-wavelength (MWL) data of the Crab Nebula, using a model that features radially dependent electron, seed photon, and magnetic field densities, finding that they can describe the entire data set well.

In this work, we provide, for the first time, a simultaneous measurement of the energy spectrum and spatial extension of the entire IC component of the Crab Nebula.
To this end, we combined data from \fermi and \hess at the level of detected \gam-ray events in a joint likelihood fit, using the open-source software package \gp \citep{Donath2023}.
In addition, we fitted phenomenological models based on those developed by \citet{Kennel1984a}, \citet{Meyer2010}, and \citet{Dirson2023} to the \fermi, \hess, and available synchrotron MWL data.
We emphasise that in contrast to previous works, our analysis does not rely on fitting models to derived data products such as flux points, which depend on the assumed underlying spectral or spatial models used in their derivation.

The remainder of the paper is structured as follows.
In Sect.~\ref{sec:data_analysis}, we first present separate analyses of the \hess and \fermi data.
The joint analysis detailed in Sect.~\ref{sec:combined_analysis} then provides a measurement of the energy spectrum and extension of the Crab Nebula across the energy range covered by both instruments.
In Sect.~\ref{sec:modelling}, we describe the fit of several phenomenological models to our combined data set. 
The impact of systematic uncertainties is discussed in Sect.~\ref{sec:systematics}.
Finally, we conclude in Sect.~\ref{sec:conclusion}.

\section{Data analysis}
\label{sec:data_analysis}
In parallel with combining the \fermi and \hess data we performed separate analyses using the standard analysis tools of the respective instruments.
These results are used to cross-check the joint analysis, which we carried out with the \gp package \citep[v0.18.2;][]{Donath2023,Deil2020}.

\subsection{Likelihood method}
\label{sec:llh_method}
The analysis is performed as a three-dimensional (longitude, latitude, and energy), binned likelihood fit.
The measured events are binned into a `counts' cube according to their reconstructed energy and sky direction.
The spectral and spatial models are evaluated on a similar cube, which can have a different binning or larger dimensions for better accuracy.
The model prediction is then forward-folded with the instrument response functions (IRFs: effective area, point spread function (PSF), and energy dispersion) in order to compute the number of predicted counts for each cell of the counts cube.
The likelihood $\mathcal{L}$ is a measure of how well the model prediction agrees with the measurement and can be calculated as
\begin{linenomath*}
\begin{equation}
    \mathcal{L}(\xi) = \prod_i \mathcal{P}(n_i | \nu_i(\xi)) = \prod_i \frac{\nu_i^{n_i}(\xi)}{n_i!} \cdot \exp(-\nu_i(\xi))\,,
\end{equation}
\end{linenomath*}
with $\mathcal{P}(n_i | \nu_i(\xi))$ the Poisson probability to measure $n_i$ events in cell $i$ given the model prediction $\nu_i$ for a set of parameters $\xi$.
Besides \gam rays from the source under investigation, there are also contributions to $n_i$ and $\nu_i$ from background processes, as will be described in more detail in the subsequent sections concerning the different instruments.
By minimising the `Cash statistic' \citep{Cash1979}, defined as
\begin{linenomath*}
\begin{equation}
    \mathcal{C} \equiv -2\ln\mathcal{L} = 2\sum_i (\nu_i - n_i\ln(\nu_i) + \ln(n_i!))
\end{equation}
\end{linenomath*}
(where the term $\ln(n_i!)$, being independent of the model parameters $\xi$, is usually omitted), the fit maximises the likelihood and thus the agreement between model and data.

Because the synchrotron emission of the Crab Nebula measured in the X-ray regime is steady to within $\pm 5\%$ over the period considered here \citep{Jourdain2020}, we do not expect variations of the flux level or spatial appearance of the IC emission.
We therefore do not take into account the time of data taking in our analysis, but perform a time-independent modelling.

\subsection{\hess data analysis}
\label{sec:hess_data_analysis}
\hess is an array of five imaging atmospheric Cherenkov telescopes (IACTs) located in Namibia at an altitude of \SI{1800}{m} \citep{HESS2006}.
It is sensitive to \gam rays in the energy range from several tens of~GeV to $\sim$\SI{100}{TeV}, where the exact achievable energy threshold depends on the observing conditions (in particular the zenith angle).
The initial array, denoted \hess Phase-I, was completed in 2003 and comprises four telescopes (CT1-4) with \SI{108}{\square\meter} mirror area each, arranged in a square grid.
We refer to data taken with this configuration as `stereo' data.
In 2012, a fifth, larger telescope (CT5) with \SI{614}{\square\meter} mirror area was added in the centre of the array, marking the beginning of \hess Phase-\RNum{2}.
The larger mirror area of CT5 can in favourable conditions lead to energy thresholds significantly below \SI{100}{GeV} \citep[e.g.][]{HESS2017_AGNmono,HESS2018_VelaPSR}.
When analysed separately, data taken with the CT5 telescope is referred to as `mono' data \citep{Murach2015}.
In 2019, the camera of CT5 was replaced by a `FlashCam' prototype camera \citep{Bi2021,Puehlhofer2021} that has been designed for the upcoming Cherenkov Telescope Array \citep[CTA;][]{CTA2018}.
\hess data taking is conducted in observation runs that typically last \SI{28}{\minute}.

IACTs are built to measure the Cherenkov light emitted in \gam-ray induced particle showers in the atmosphere.
Based on the recorded shower images, we reconstructed the direction and energy of the primary \gam ray using the \textsc{ImPACT} reconstruction algorithm \citep{Parsons2014}.
The dominant background consists of air showers initiated by charged cosmic rays, which we suppress by means of a multi-variate analysis \citep{Ohm2009}.
In contrast to traditional methods that estimate the residual background from `OFF' regions in the same observation run, we used here a three-dimensional background model constructed from archival \hess observations \citep[this is a prerequisite for carrying out the three-dimensional likelihood analysis described at the beginning of this section;][]{Mohrmann2019}.
The IRFs are obtained from extensive Monte-Carlo simulations, carried out with the \textsc{sim\_telarray} package \citep{Bernloehr2008}.
A correction for telescope efficiency, measured for each observation run using muons from atmospheric air showers, is applied as described in \citet{Mitchell2016}.
Results have been cross-checked with independent calibration and reconstruction methods \citep{deNaurois2009}.

We used 114 observation runs (corresponding to an observation time of $\approx$50 hours) of \hess stereo (CT1-4) data taken between Nov.\ 2004 and Mar.\ 2015.
Only observations with good data quality, excellent atmospheric conditions, zenith angles below 55$^\circ$, and a maximum angle between the pointing position and the Crab Nebula of 1$^\circ$ have been included.
Compared to the standard analysis configuration, we required a minimum of three telescopes for the reconstruction of each event, which leads to fewer detected events mostly at low energies but improves the angular resolution by $\sim$20\%, to $0.06^\circ$ (68\% PSF containment radius) at \SI{1}{TeV}.
At the same energy, the achieved energy resolution is 16\% (standard deviation).
Because the Crab Nebula culminates at a relatively large zenith angle of $\sim$$45^\circ$ at the \hess site, the energy threshold of the stereo data set is comparatively high ($\approx$\SI{560}{GeV}).

\begin{table}
    \centering
    \caption{Overview of \hess data.}
    \label{tab:hess_data_sets}
    \begin{tabular}{ccS[table-number-alignment=right]S[table-number-alignment=right]}
        \hline\hline
        {Data set} & {Year} & {Runs} & {Observation time (h)}\\
        \hline
        stereo & 2004 & 18 & 7.82\\
               & 2005 & 6 & 2.81\\
               & 2007 & 5 & 2.34\\
               & 2008 & 4 & 1.43\\
               & 2009 & 10 & 4.69\\
               & 2013 & 58 & 25.19\\
               & 2014 & 7 & 3.01\\
               & 2015 & 6 & 2.80\\
        \hline
        mono & 2019 & 18 & 8.41\\
             & 2020 & 42 & 19.44\\
             & 2021 & 5 & 2.09\\
        \hline
    \end{tabular}
    \tablefoot{
        The separation into years is done only for illustrative purposes here; the data are not separated by year when they enter the analysis.
    }
\end{table}

Additionally, we used 65 observation runs ($\approx$30 hours) of mono data taken with the CT5 telescope with its new camera, between Nov.\ 2019 and Oct.\ 2021.
This data set has a lower energy threshold ($\approx$\SI{240}{GeV}) compared to the stereo one, which increases the overlap with the energy range covered by the \Fermi (see next section).
Because the reconstruction of events uses information from only one telescope in this case, the performance in terms of angular resolution ($0.13^\circ$) and energy resolution (29\%) is worse compared to the stereo data set.
A summary of the \hess data sets can be found in Table~\ref{tab:hess_data_sets}.

The region of interest (ROI) for the \hess data analysis is defined as a region of $5^\circ\times 5^\circ$ around the position of the Crab Nebula, in Galactic coordinates.
Because the extension of the Crab Nebula is at the limit of detectability for \hess, we chose a relatively fine spatial bin size of $0.01^\circ$, to achieve an accurate description of the PSF and to not smear the signal artificially.
Furthermore, we binned the data in energy using 16 bins per decade.
For better performance, we only considered events with a maximum offset of 1.5$^\circ$ from the pointing position in each observation run.

In a first step, we fitted the normalisation and spectral tilt\footnote{The spectral tilt parameter $\delta$ modifies the predicted background rate $R$ at energy $E$ as $R^{\prime}=R\cdot(E/\SI{1}{TeV})^{-\delta}$.} of the background model template to each observation individually, where we excluded a circular region with radius $0.3^\circ$ around the Crab Nebula.
For the stereo data, we then binned the observations by zenith angle into three bins with edges $[45, 47.5, 52.5, 55]^\circ$, so that the observations in each bin have nearly identical energy thresholds.
The mono observations exhibit a smaller variation in zenith, so that this procedure is not necessary.
For each bin, as well as for the mono data, we combined the respective observations to a stacked data set.
In the stacking process, the counts, exposure, and predicted background are summed while the energy dispersion and point spread function are averaged.
The stacking makes the analysis computationally less expensive at the cost of using slightly less accurate IRFs.
The binning of observations in zenith angle prevents the averaging of IRFs across observation runs with too dissimilar observing conditions.
We additionally smoothed the runs-averaged PSF with a $30''$ one-dimensional Gaussian kernel, which represents the scale of pointing uncertainty of the telescopes \citep{Braun2007}, where we assume that the pointing error varies from run to run.
After this procedure, we are left with three stereo data sets (for low, medium, and high zenith angles) as well as one mono data set.

Subsequently, we modelled the \gam-ray emission of the Crab Nebula, that is, we performed a three-dimensional likelihood analysis as introduced in Sect.~\ref{sec:llh_method} to simultaneously fit the spatial morphology and the energy spectrum.
We used a symmetric, two-dimensional Gaussian as spatial model for the intrinsic source distribution, which is then folded with the PSF.
At this stage, we did not yet consider a possible energy dependence of the spatial model.
All extension measurements reported in this work denote the 68\% containment radius of the emission, which for a two-dimensional Gaussian with width $\sigma$ is given by $r_{68} = \sqrt{-2\cdot\ln(1-0.68)}~\sigma \approx 1.5\cdot\sigma$.
From the stereo data set, we obtain a fitted extension (averaged over energy) of $1.62'\pm 0.05'_\mathrm{stat}~{}^{+0.21}_{-0.24}{}'_\mathrm{sys}$.
The systematic uncertainty is derived by repeating the analysis with a PSF that has been broadened (narrowed) by 5\% with respect to the default one, independent of energy.
The variation of 5\% in width has been established as a realistic value in a study -- with the same analysis configuration as employed here -- of the bright blazar PKS~2155$-$304 \citep[see e.g.][]{HESS_PKS2155_2010}, which appears as a point-like source for the \hess array.
It amply covers the $30''$-smoothing applied to the PSF earlier.
Finally, we note that given the larger systematic uncertainty and considerably larger PSF of the mono data, we cannot constrain the extension with this data set.

The obtained value is larger than the one found in \citet{HESS2020}, $r_\mathrm{68}=1.30'\pm0.07'_\mathrm{stat}\pm 0.17'_\mathrm{sys}$, although we note that the deviation is far from significant when taking into account the systematic uncertainties.
We attribute the difference partly to the fact that at the lower energy threshold achieved here (\SI{560}{GeV} instead of \SI{700}{GeV}), the extension is indeed expected to be slightly larger.
Nevertheless, we note that this comparison of results indicates that the systematic uncertainty may be slightly underestimated.
The measured extension is compatible with the upper limits provided by the HEGRA telescopes \citep{HEGRA2000,HEGRA2004}.

\begin{figure}
    \centering
    \includegraphics[width=0.99\linewidth]{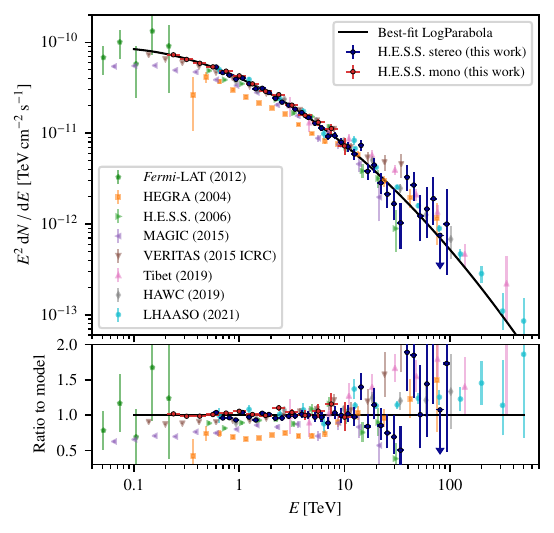}
    \caption{
        \hess flux points of the mono (CT5) analysis (red) and stereo (CT1-4) analysis (blue) in comparison to results from other instruments.
        The error bars show the 68\% statistical uncertainty.
        Points less significant than 2$\sigma$ are shown as 95\% confidence level upper limits.
        The black line displays the result of fitting a log-parabola model to the \hess data of this work.
        References (in the order of the legend): \citet{HEGRA2004}, \citet{HESS2006}, \citet{MAGIC2015}, \citet{VERITAS2015}, \citet{Tibet2019}, \citet{HAWC2019}, and \citet{LHAASO2021}.
    }
    \label{fig:hess_sed}
\end{figure}

For the spectral model, we used a log-parabola function,
\begin{linenomath*}
\begin{equation}
\label{eq:log-parabola}
    \frac{dN}{dE} = N_0 \left(\frac{E}{E_0}\right)^{-\alpha - \beta\ln(E/E_0)},
\end{equation}
\end{linenomath*}
with normalisation~$N_0$, reference energy~$E_0$, spectral index~$\alpha$, and curvature parameter~$\beta$.
From a joint fit to all \hess data sets, we obtain $N_0=(4.06\pm0.03)\times 10^{-11}\,\mathrm{TeV}^{-1}\,\mathrm{cm}^{-2}\,\mathrm{s}^{-1}$ at $E_0=1\,\mathrm{TeV}$ (fixed), $\alpha=2.530\pm0.006$, and $\beta=0.086\pm0.005$; this result is shown as a black line in Fig.~\ref{fig:hess_sed}.
For both the mono and stereo data, we subsequently derived separate sets of flux points by first performing the fit separately for each data set, and then re-adjusting the normalisation parameter in narrow energy ranges (where we have ensured that the best-fit models for the two sets are compatible with each other).
Taking into account the relatively poor energy resolution of the mono data, we reduced the number of flux points (from 16 to 8 per decade in energy) for this data set in order to avoid correlations between the points.
Both the stereo and mono flux points are also displayed in Fig.~\ref{fig:hess_sed}.
The figure illustrates that the mono data set allows us to extend the measurement to lower energies, whereas the stereo data set extends up to almost $\sim$\SI{100}{TeV}.
A comparison with results from other instruments shows that our derived flux is largely compatible, with the noticeable deviations most likely explained by a systematic shift in the energy scale of the instruments.
The spectral results have been confirmed in several one-dimensional analyses based on the more traditional reflected-background method \citep{Fomin1994,Berge2007}.
One example of such a comparison is shown in Appendix~\ref{appx-comp-hess}.
Finally, we note that in contrast to the \fermi analysis (see next section), the \gam-ray emission from the Crab Pulsar itself, despite extending into the TeV energy range \citep{MAGIC_CrabPulsar_2016}, can safely be neglected for our study.
Adopting the measured spectrum of the pulsed emission from \citet{MAGIC_CrabPulsar_2016}, even the contamination for the lowest-energy flux point derived with the \hess mono data set is below 0.5\%.

\subsection{\fermi data analysis}
\label{sec:fermi_data_analysis}
The \Fermi is a pair conversion telescope designed to measure \gam~rays with energies from \SI{20}{MeV} to above \SI{300}{GeV}~\citep{FermiLAT_Detector_2009}.
We first performed a pulsar phase-resolved reference analysis of the Crab Nebula using the standard \textsc{fermitools} version 1.2.23\footnote{\url{http://fermi.gsfc.nasa.gov/ssc/data/analysis/software}} and \textsc{fermipy}, version 0.19.0+14\footnote{\url{http://fermipy.readthedocs.io}}~\citep{Wood2017}.
This analysis serves as a comparison for the \gp analysis. 

\begin{figure}
    \centering
    \includegraphics[width=.9\linewidth]{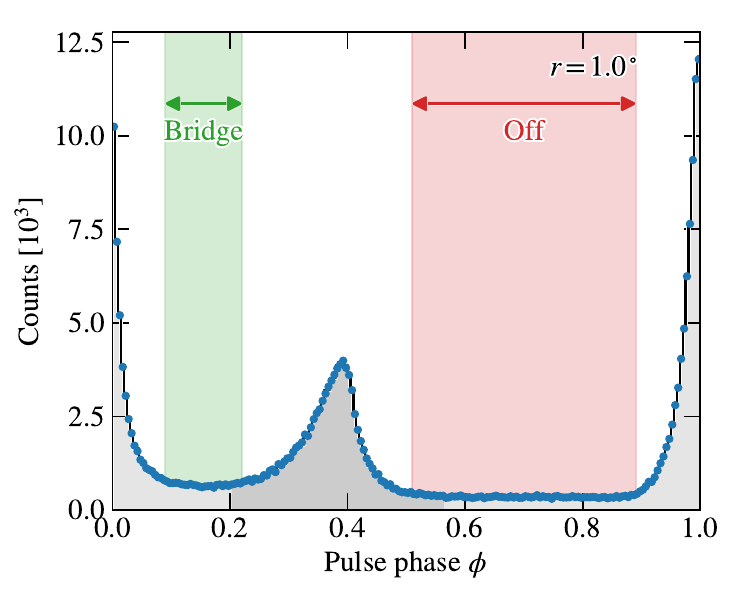}
    \caption{
        Distribution of pulsar phases of \gam-ray candidate events detected with \fermi around the Crab Nebula.
        The distribution includes events with a reconstructed arrival direction that is within $1^\circ$ of the Crab Nebula's position.
        The shaded regions indicate the intervals for the Bridge emission ($0.09\leqslant\phi\leqslant 0.22$, not relevant here) and Off region ($0.51\leqslant \phi \leqslant 0.89$) derived with the QB algorithm.
        The grey shaded areas indicate regions found by the HOP algorithm, which identifies `watersheds' between local maxima~\citep{Meyer2019}.
        There are shown here for illustration only.
    }
    \label{fig:pulses}
\end{figure}

To this end, we selected \gam~rays that were observed between 4~August 2008 and 4~January 2020 in the energy range between \SI{100}{MeV} and \SI{3}{TeV} and within a $10^\circ\times10^\circ$ ROI centred on the nominal Crab Nebula position provided in the fourth \fermi \gam-ray source catalogue \citep[4FGL;][]{FermiLAT_4FGL_2020}.
As we are interested in emission of the nebula and not the pulsar, we proceeded to determine the off-pulse phase interval.
To do so, we used an ephemeris prepared following the methods described in \citet{FermiLAT_Ephemeris_2022} and generated a histogram of the pulsar phases $\phi$ of all photons that have arrived within a $1^\circ$ radius around the Crab Nebula, see Fig.~\ref{fig:pulses}.
The off-pulse phase interval is determined with the `quiescent background' (QB) algorithm introduced by \citet{Meyer2019}.
The resulting interval is $0.51\leqslant \phi \leqslant 0.89$, as indicated by the red-shaded region in Fig.~\ref{fig:pulses}.

In the 4FGL, the total emission from the Crab Nebula is modelled with two super-imposed source models for the high-energy tail of the synchrotron emission and the emission due to IC scattering, respectively.
For the combination with \hess data, we are mostly interested in the IC part of the spectrum.
As the synchrotron and IC components are of similar intensity around \SI{1}{GeV}, we proceeded in an iterative fashion: we first optimised an ROI model over the entire energy range.
We then fixed the synchrotron component to the best-fit model and performed a second analysis of the nebula above \SI{1}{GeV}, which is dominated by IC radiation.
For this second analysis, we also used a finer spatial binning (see below) in order to be able to determine the extension of the IC nebula. 

For the initial ROI model, we selected photons within the off-pulse interval and we further excluded events that have arrived at a zenith angle above $90^\circ$ in order to mitigate contamination of \gam~rays originating from Earth's limb. 
We also excised periods of bright \gam-ray bursts and solar flares that have been detected with a test statistic $(\mathrm{TS}) > 100$.\footnote{The TS is defined as $\mathrm{TS} = -2\ln(\mathcal{L}_1 / \mathcal{L}_0)$, i.e.\ the log-likelihood ratio between the maximised likelihoods $\mathcal{L}_1$ and $\mathcal{L}_0$ for the hypotheses with and without an additional source, respectively~\citep{Mattox1996}.}
We used the \texttt{Pass 8} IRFs~\citep{FermiLAT_Pass8_2013} and selected \gam~rays passing the \texttt{P8R3 SOURCE} event selection.
We chose a spatial binning of $0.04^\circ\,\mathrm{pixel}^{-1}$ and eight energy bins per decade.

\begin{figure}
    \centering
    \includegraphics[width=.98\linewidth]{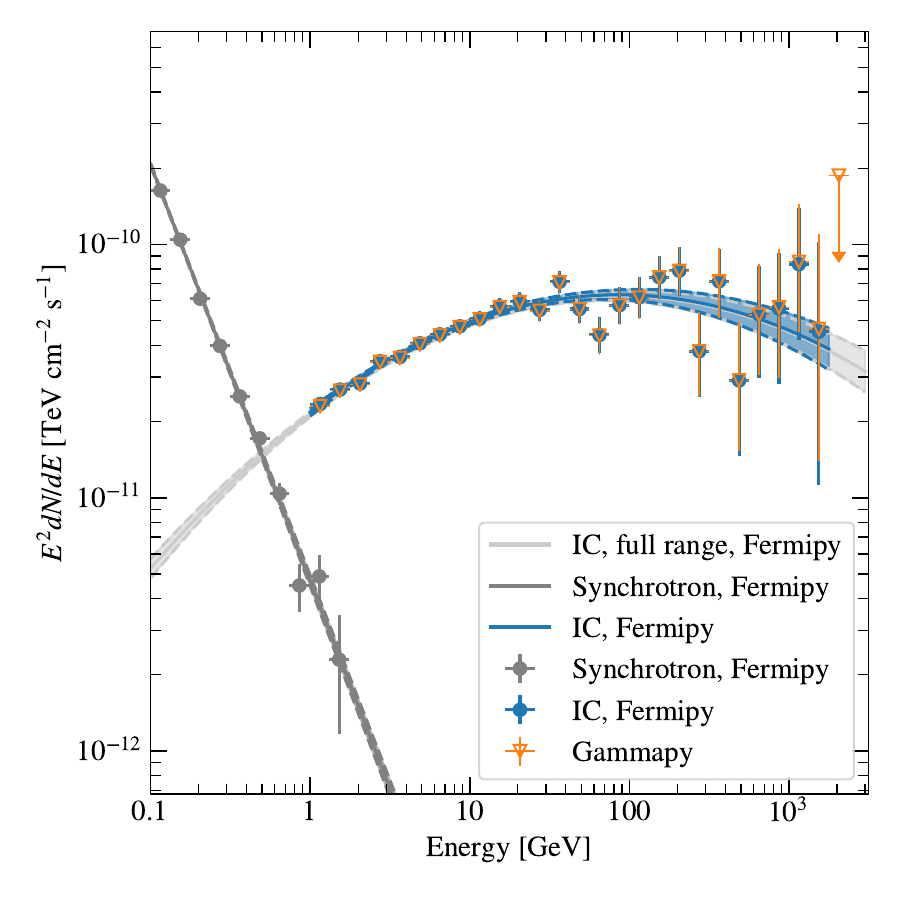}
    \caption{
        \fermi spectra of synchrotron and IC emission.
        The IC emission is derived with both \fermipy and \gp and the spectra agree remarkably well.
        Details on the \gp analysis of \fermi data are given in Sec.~\ref{sec:prep_fermi_data}.
        The error bars on the flux points show the 68\% statistical uncertainties.
     }
    \label{fig:fermi_sed}
\end{figure}

We initialised the ROI model with all point-like and extended \gam-ray sources within $20^\circ$ of the ROI centre as listed in the 4FGL (except the Crab Pulsar), from which we also took the initial spectral source parameters.
Additionally, the standard templates for isotropic and Galactic diffuse emission are used.\footnote{We used the file gll\_iem\_v07.fits or the Galactic diffuse emission and the file iso\_P8R2\_SOURCE\_V6\_v07.txt for the isotropic diffuse component; see: \url{ http://fermi.gsfc.nasa.gov/ssc/data/access/lat/BackgroundModels.html}}
In the \fermi energy range, the synchrotron component is expected to exhibit only a very small spatial extension that is not resolvable with \fermi, and therefore modelled as a point-like source.
For the IC component, we used the 4FGL default source morphology, which is given by a radial Gaussian with a width of $\sigma = 0.0198^\circ$ ($r_{68}=1.79'$).
As for the spectrum, the synchrotron part is modelled with a simple power law, 
\begin{linenomath*}
\begin{equation}
    \frac{dN}{dE} = N_0 \left(\frac{E}{E_0}\right)^{-\Gamma},
\end{equation}
\end{linenomath*}
with normalisation $N_0$, reference energy $E_0$, and spectral index $\Gamma$, whereas the IC component is described with a log-parabola as defined in Eq.~\ref{eq:log-parabola}.
After an initial optimisation, we freed the spectral normalisation and spectral shape parameters of sources within $10^\circ$ from the ROI centre as well as all spectral parameters of the isotropic and diffuse emission templates.
We froze all spectral parameters for sources with $\mathrm{TS} < 5$.
After successful convergence of the fit, we generated a TS map to search for additional point sources.
For each pixel in the ROI, we added a putative point source with a power-law spectrum with index $\Gamma = 2$ and calculated its TS.
No additional sources with $\sqrt{\mathrm{TS}} \geqslant 5$ were found.
The best-fit spectrum of the synchrotron component is shown in Fig.~\ref{fig:fermi_sed} and the parameters are provided in Table~\ref{tab:fermi_sed}.

The best-fit ROI model was then used to perform an analysis above \SI{1}{GeV}.
In terms of data selection, we followed the official \fermi recommendations and relaxed the zenith angle cut from $90^\circ$ to $100^\circ$.
We shrank the ROI to $6^\circ\times6^\circ$ and following \citet{FermiLAT_ExtendedSources_2018}, we chose a finer spatial binning of $0.025^\circ\,\mathrm{pixel}^{-1}$ that we apply in the calculation of the livetime and exposure cubes.
Furthermore, we made use of PSF event types \citep{FermiLAT_Detector_2009}.
The data are split into four sets according to the quality of the reconstruction of the \gam-ray arrival directions.
Each set is analysed with its own specific IRFs and the four sets are combined by multiplying their respective likelihood values.
Accordingly, the isotropic background templates for these event types have been used. 
We also limited the maximum energy of the analysis to the edge of the energy bin that contains the highest energy photon, that is, \SI{1.78}{TeV}.
The spectral parameters of the synchrotron component have been fixed.
The resulting spectrum is presented in Fig.~\ref{fig:fermi_sed} and the best-fit spectral parameters are listed in Table~\ref{tab:fermi_sed}.
They agree well with the spectral parameters of the fit over the whole energy range.

\begin{table*}
    \centering
    \caption{Best-fit parameters of synchrotron and IC emission models of the Crab Nebula derived from \fermi data.}
    \label{tab:fermi_sed}
    \begin{tabular}{lcccc}
        \hline\hline
         & $N_0$ & $E_0$ & $\Gamma$ or $\alpha$ & $\beta$\\
         & ($10^{-7}$ MeV$^{-1}$ cm$^{-2}$ s$^{-1}$) & (MeV) & & \\
         \hline
         Synchrotron &  2.47 $\pm$ 0.06 & 50.6 & 3.638 $\pm$ 0.024 & -- \\
         IC (full energy range) & $(4.75 \pm 0.10)\times10^{-6}$ & 
          $10^{4}$ & 1.762 $\pm$ 0.012 & 0.053 $\pm$  0.006 \\ 
        IC ($E \geqslant$ 1\,GeV) & $(4.93 \pm 0.11)\times10^{-6}$
         & $10^{4}$ & 1.768 $\pm$ 0.013 & 0.054 $\pm$  0.008 \\ 
        IC ($E \geqslant$ 1\,GeV, \gp) & $(4.88 \pm 0.11)\times10^{-6}$
         & $10^{4}$ & 1.763 $\pm$ 0.013 & 0.056 $\pm$ 0.008 \\
        \hline
    \end{tabular}
\end{table*}

Finally, we obtained the extension of the nebula using a two-dimensional Gaussian model as before.
In this procedure, the nebula's sky coordinates are additional free parameters, as are the spectral normalisation parameters of the diffuse background sources.
The spectral parameters of the nebula itself are frozen to their best-fit values.
We obtain a 68\,\% containment radius of $r_\mathrm{68}=2.22'\pm0.18'_\mathrm{stat}~{}^{+0.54}_{-0.48}{}'_\mathrm{sys}$, where the systematic uncertainties are derived by using a bracketing method for the PSF~\citep{FermiLAT_ExtendedSources_2018}.
We find a TS value for the extension of TS$_\mathrm{ext}=47.6$ with the nominal PSF, which reduces to TS$_\mathrm{ext}=16.3$ if the larger bracketing PSF is used instead.
These results are compatible with the extension found by \citet{FermiLAT_ExtendedSources_2018}, $r_\mathrm{68}=1.80'\pm0.18'_\mathrm{stat}\pm 0.42'_\mathrm{sys}$, within the uncertainties.

\section{Combined analysis}
\label{sec:combined_analysis}
In contrast to previous analyses, in which final data products such as flux points or extension measurements from different instruments have been analysed together, here we combine the \fermi and \hess data at the event level.
The advantage of this approach is that we are able to exploit more information and ensure consistency of spectral and morphological models between the different instruments.
In the combined analysis we simultaneously fitted one three-dimensional source model (spectrum and morphology) to the (binned) event data of \fermi and \hess.
The analysis is carried out with \gp, where we used the data sets which we generated and cross-checked for the individual analyses and summed their respective Cash statistic values in the combined analysis.
The total $\mathcal{C}$ value is then minimised, which corresponds to a joint fit of the model to the \fermi and \hess data sets.

\subsection{Preparation of \fermi data in \gp}
\label{sec:prep_fermi_data}
During the \fermipy analysis the events are binned into a counts cube, and the IRFs (exposure, energy dispersion, and PSF) are calculated.
We used the data products generated during the \fermi analysis to set up the combined analysis within \gp, where we generated four separate \fermi data sets, corresponding to the four PSF classes.
While the analyses with \fermipy and \gp are in principle very similar, there are some differences which require adaptations in the data production steps.
These are described in the following.

First, \gp only calculates model fluxes of sources within the ROI, whereas the standard \fermi analysis tools allow the inclusion of source models outside the ROI, which may contribute to the observed counts in the ROI especially at low energies, where the \fermi PSF is very broad.
Consequently, the ROI size of the \fermi data sets has been increased to $10^\circ\times10^\circ$, such that the 4FGL sources outside the inner ($6^\circ\times6^\circ$) analysis region are also evaluated by \gp.
Additionally, to correctly take into account the energy dispersion at the boundaries of the analysed `reconstructed' energy range with \gp, we need to be able to evaluate the IRFs at `true' \gam-ray energies that extend beyond this range.
As it is not possible to separately define the ranges to be considered for reconstructed and true energy in \fermipy, we increased the energy range by two bins at the low-energy end and by one bin at the high-energy end.
The added spatial and energy bins were then again masked in the \gp analysis, so that exactly the same regions and energy bins as in the \fermipy analysis contribute to the likelihood calculation.

To verify that our analysis setup in \gp is correct, we repeated the analysis of the \fermi data previously carried out with \fermipy (cf.\ Sect.~\ref{sec:fermi_data_analysis}), using exactly the same model components.
In the fit, the model parameters of the Crab Nebula's IC emission as well as normalisation of the isotropic diffuse and Galactic diffuse backgrounds and spectral index of the Galactic diffuse model are left free, while all parameters of the other source models in the ROI are fixed to the best-fit values from the \fermipy analysis.
With this procedure, we obtain an excellent agreement between the \fermipy and the \gp analysis.
For the spectrum, this is illustrated in Fig.~\ref{fig:fermi_sed} and summarised in Table~\ref{tab:fermi_sed}.
For the extension, the \gp fit yields $r_{68}=(2.10\pm 0.18_\mathrm{stat})'$, which is consistent with the \fermipy result given at the end of Sect.~\ref{sec:fermi_data_analysis}.

\subsection{Measuring the Crab Nebula's combined energy spectrum and extension}
\label{sec:ext_measurement}
\begin{figure}
    \centering
    \includegraphics[width=.98\linewidth]{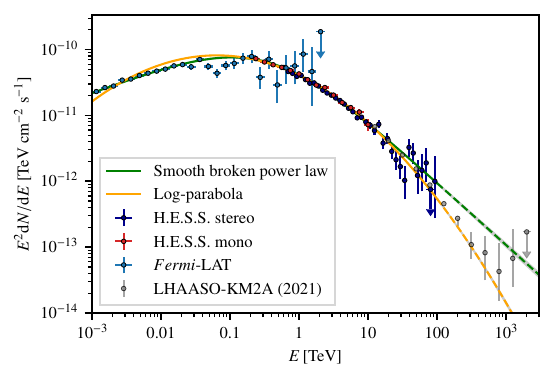}
    \caption{
        Comparison of log-parabola and smoothly broken power law spectral models.
        Both are fitted to the \fermi and \hess data sets, while the LHAASO flux points are not included in the fit and only shown for comparison.
     }
    \label{fig:lp-vs-sbpl}
\end{figure}

Through the combined fit, we can now consistently measure the extension of the Crab Nebula across the combined energy range of \fermi and \hess.
To this end, we fitted our models jointly to the \fermi and \hess data sets as defined in the previous sections.
As in the separate analyses, we used a two-dimensional Gaussian as the spatial model.
Because the log-parabola model does not yield a satisfactory description of the IC component across the entire energy range covered by our combined fit, we used a smoothly broken power-law model for the spectral component:
\begin{linenomath*}
\begin{equation}
    \frac{dN}{dE} = N_0 \left(\frac{E}{E_0}\right)^{-\Gamma_1}\left[1+\left(\frac{E}{E_\mathrm{break}}\right)^{\frac{\Gamma_2 - \Gamma _1}{\beta}}\right]^{-\beta}\,,
\end{equation}
\end{linenomath*}
with $E_0$ fixed at \SI{1}{TeV}.
A fit over the whole energy range yields the following parameters: $N_0=(3.35 \pm 0.22)\times 10^{-10}\,\mathrm{TeV}^{-1}\mathrm{s}^{-1}\mathrm{cm}^{-2}$, $\Gamma_1 = 1.61\pm 0.02$, $\Gamma_2 = 2.95\pm 0.02$, $E_\mathrm{break} = 0.33\pm 0.02\,\mathrm{TeV}$, and $\beta = 1.73\pm 0.07$.
This model is preferred over the log-parabola model by a difference of~$94$ in the Akaike information criterion, $\mathrm{AIC}=\mathcal{C}+2N_\mathrm{par}$ \citep{Akaike1974}, where the number of free model parameters is $N_\mathrm{par}=3$ for the log-parabola model and $N_\mathrm{par}=5$ for the smoothly broken power-law model.
Both models are shown in Fig.~\ref{fig:lp-vs-sbpl}, together with the \fermi and \hess flux points derived in this work.
A comparison of the models with the LHAASO flux points, also shown in the plot, indicates that an extension of the analysis beyond \SI{100}{TeV} would require the addition of another break to the spectral model.

The best-fit position over the whole IC energy range is $(184.5499\pm 0.0004_\mathrm{stat})^\circ$ Galactic longitude and $(-5.7825\pm 0.0004_\mathrm{stat})^\circ$ Galactic latitude.
This position is shifted to the north with respect to that of the Crab Pulsar by $\approx 28''$.
In what follows, the only parameter of interest is the extension parameter of the Gaussian spatial model.
However, in order to take into account possible correlations with other parameters, the following parameters are also left free: amplitude and position of the Crab Nebula source model, as well as the normalisation parameters of the \fermi isotropic and Galactic diffuse background models and the \hess background models. 
The rest of the spectral parameters as well as the rest of the \fermi ROI model remain fixed.
To obtain an energy-dependent measurement of the extension, we optimised the free parameters separately in different energy intervals, where we used a binning of two bins per decade in the observed energy range between \SI{1}{GeV} and \SI{100}{TeV}.
As an important cross-check, we also compared the best-fit position of the Crab Nebula spatial model between all energy bands, finding that they are all compatible with each other within the uncertainties.
Furthermore, the fitted extensions remain unchanged if the centre of the spatial model is fixed to a common position for all energy bands.

As the PSF of the \hess mono data set is not understood well enough, we did not include this set in the measurement of the extension.
Potentially large systematic effects may otherwise dominate the extension measurement between \SI{240}{GeV} and \SI{560}{GeV} (where the stereo data set has its threshold), due to the large number of events detected at these energies.

\begin{figure}
    \centering
	\includegraphics[width=0.99\linewidth]{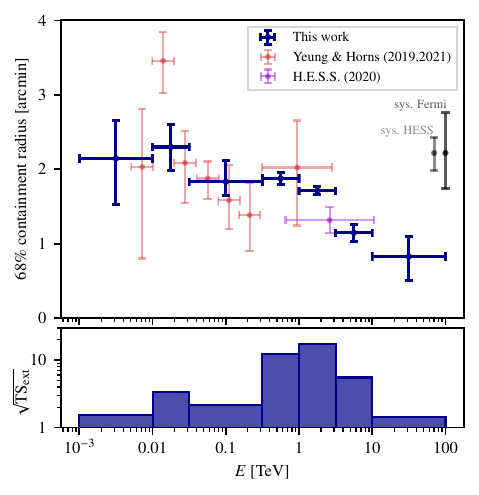}
	\caption{
        68\% containment radius of the \gam-ray emission from the Crab Nebula measured with the combined \fermi and \hess data analysis.
        The error bars denote the 1$\sigma$ statistical uncertainties and do not include systematic effects.
        For comparison, the red points show the extension measured with \fermi by \citet{Yeung2019,Yeung2021}, whereas the purple point displays the extension previously measured with \hess \citep{HESS2020}.
        In grey we show the systematic uncertainty of the \hess and \fermi measurements, respectively (cf. Sect.~\ref{sec:hess_data_analysis} \& Sect.~\ref{sec:fermi_data_analysis}).
        The bottom panel shows the square root of the TS value of the measured extension, which gives an indication of the (statistical) significance of the extension with respect to the null hypothesis of a point-like source.
    }
	\label{fig:extensions-raw}
\end{figure}

In Fig.~\ref{fig:extensions-raw} we show the measured extension as a function of the \gam-ray energy.
Below the threshold of the \hess stereo data set, \SI{560}{GeV}, only the \fermi data contribute.
To obtain a more significant extension, we have merged the first and last two bins in this range.
Above \SI{560}{GeV}, the \hess data immediately dominate the fit due to the much larger effective area and better angular resolution.
The highest significance for an extension is reached in the energy bin ranging from \SIrange{1}{3}{TeV}; this range also dominates the energy-integrated extension measurement with \hess presented in Sect.~\ref{sec:hess_data_analysis}.

The extensions measured in the ranges dominated by the \fermi and \hess data, respectively, connect smoothly and are furthermore consistent with published results.
Overall, we find strong evidence for an extension that decreases with energy.
As we subsequently see in Sect.~\ref{sec:modelling}, this result is in very good agreement with theoretical expectations.

\begin{figure}
    \centering
	\includegraphics[width=0.99\linewidth]{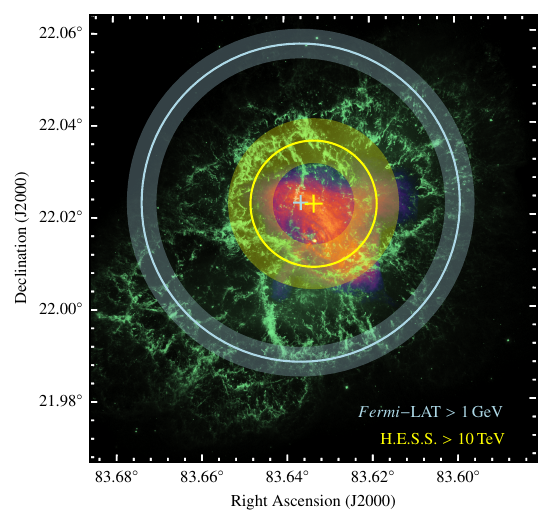}
	\caption{
        Optical image of the Crab Nebula in green (credit: NASA, ESA/Hubble), overlaid with an X-ray image in red, taken with \textit{Chandra} (\SIrange{0.3}{10}{keV}; credit: NASA/CXC/SAO, \url{https://chandra.harvard.edu/photo/openFITS/crab.html}).
        The circles show the 68\% containment radii of the \gam-ray emission measured with \fermi ($>$\SI{1}{GeV}; blue) and \hess ($>$\SI{10}{TeV}; yellow), where the bands illustrate the statistical uncertainties.
        The crosses indicate the fitted position and its uncertainty for the respective data sets.
    }
	\label{fig:extensions-im}
\end{figure}

The decreasing extension is also illustrated in Fig.~\ref{fig:extensions-im}, where we show a measurement of the Crab Nebula in the optical regime and at X-ray energies, and compare this to our extension measurements with \fermi and \hess.
As noted earlier, the fit of the \fermi data above \SI{1}{GeV} yields an extension of $r_{68}=(2.10\pm 0.18_\mathrm{stat})'$.
The extension measured with \hess above \SI{10}{TeV}, on the other hand, is $r_{68}=(0.82\pm 0.29_\mathrm{stat})'$.
In most models, as will be further detailed in the subsequent section, the optical radiation is produced by electrons that are also responsible for the IC emission in the \fermi energy range.
In agreement with this, the position and extension of the emission measured with \fermi is roughly consistent with the outline of the nebula seen in the optical image.
As the electrons radiating the X-ray synchroton photons also produce the highest-energy IC emission, a similar agreement between the X-ray image and the \hess measurement above \SI{10}{TeV} is expected.
However, while the centroid of the \gam-ray emission coincides well with a region of bright X-ray emission, we observe that the 68\% containment radius measured with \hess almost entirely encloses the X-ray emission, which is an unexpected result.
While the argument here is rather qualitative, this point will be addressed again in a more quantitative way in the discussion of the modelling results in Sect.~\ref{sec:ext}.
There, we confirm that the models investigated in this work fail to explain the difference between the extension in X-rays and at the highest-energy \gam~rays.
While the difference could also be due to underestimated systematic effects (which will be discussed in Sect.~\ref{sec:systematics}), it hints at an incompleteness of the models.

\section{Modelling of the Crab Nebula}
\label{sec:modelling}
In this section we introduce the phenomenological models that we investigate in this work.
We present the results of fitting them to the \fermi, \hess, and MWL data and discuss the implications.

\subsection{Description of the tested models}
In total, we tested three time-independent leptonic models that differ in the modelling of the high-energy electrons and the radial dependence of the magnetic field.
All models are radially symmetric synchrotron self-Compton (SSC) models, in which the low-energy photons are produced by synchrotron emission and the high-energy photons by IC scattering, where the synchrotron photons themselves serve as one of the targets.
We assume two electron distributions: relic `radio' electrons at lower energies and freshly injected `wind' electrons at higher energies, where the latter are believed to be accelerated at a standing shock.
The shock is commonly associated with a bright X-ray ring feature located at a distance of $r_s = (13.3 \pm 0.2)''$ from the pulsar \citep{Weisskopf2012}, which corresponds to $r_s \approx \SI{0.129(2)}{pc}$ for a distance of $d = \SI{2}{kpc}$ between Earth and the pulsar, which we take as the centre of the nebula in our radially symmetric models.
The electron density is both a function of energy and distance from the nebula centre. 
The spatial distribution of the radio electrons is assumed constant in energy since the synchrotron cooling times are larger than the age of the nebula.
These electrons could be the result of an injection during the early phases of the pulsar spin down \citep{Atoyan1999}.
For the wind electrons, the spatial distribution is allowed to change with energy.
All models assume three seed photon fields for the IC scattering: the cosmic microwave background (CMB, \SI{2.7}{K}), far-infrared emission from nearby dust (\SI{39}{K} and \SI{149}{K}; see \citealt{Dirson2023}), and the synchrotron photons.

For the description of the radial dependence of the magnetic field, 
we follow (1) \citet{Meyer2010} for a model where the magnetic field is constant throughout the nebula (\cbm); (2) \citet{Dirson2023}, where the magnetic field decreases (with adjustable rate) with increasing distance from the centre of the nebula (\vbm); and (3) \citet{Kennel1984} for a magneto-hydrodynamic (MHD) flow model (K\&C model).
The models are described in full detail in Appendix~\ref{appx-emission_model}. 
In the following, we refer to the former two models as static models in contrast to the K\&C model, which is a result of the MHD flow equations. 
All fit parameters of the models are summarised in Table~\ref{table:parameters} in Appendix~\ref{sec:appendix_fit_parameters}.

\begin{figure}
    \centering
	\includegraphics[width=0.99\linewidth]{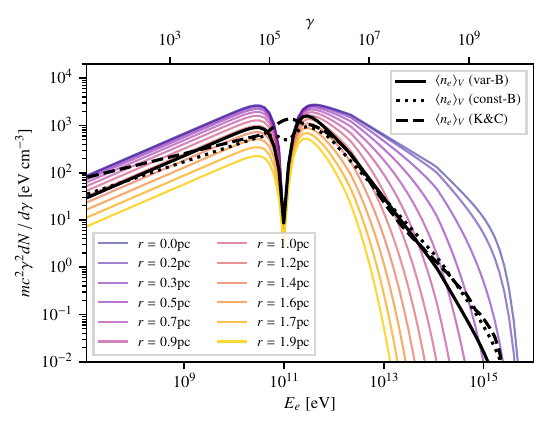}
	\caption{
        Electron densities of models, with parameters as given in Tab.~\ref{table:parameters}.
        The solid, dotted and dashed black lines show the volume-averaged density for the variable $B$-field, the constant $B$-field and the K\&C model, respectively.
        The coloured lines show the electron density for the variable $B$-field model at different radii.
        The radio electrons (at energies below the `dip') are modelled with an energy-independent extension, therefore the density drops with radius independent of energy.
        On the other hand, the density of the freshly injected wind electrons decreases more rapidly for higher energies.
	}
	\label{fig:el-spec}
\end{figure}

In all models, the radio electron density follows a simple power law in energy between the Lorentz factors $\gamma_{r,\mathrm{min}}$ and $\gamma_{r,\mathrm{max}}$, where a super-exponential cutoff occurs.
This spectral distribution is multiplied with a Gaussian to model the spatial dependency from the centre of the Crab Nebula.
For the wind electrons, the constant and variable $B$-field models assume an electron density parameterised for different energies and radii, as shown in Fig.~\ref{fig:el-spec}.
Specifically, the wind electron distribution is modelled with a double-broken power law with super-exponential cut-offs at low and high energies.
These two models are only distinguished by the spatial shape of the magnetic field, which is either constant or, for the \vbm, proportional to $r^{-\alpha}$.
In the K\&C model, on the other hand, the wind-electron spectrum is the result of an injection of electrons at the pulsar wind termination shock, which  then evolve (together with the $B$-field) according to the solution of the static MHD flow equations matching the boundary condition of flow speed and position of the shock.
In addition to the parameters of the electron injection spectrum, the model depends on the magnetisation $\sigma$ and spin-down luminosity (see Appendix~\ref{appx-emission_model} for details).

\subsection{Fitting of the models}
All three models give a prediction of the \gam-ray intensity as a function of energy and angular separation from the Crab Pulsar, which we forward-folded with the IRFs and fitted to the \fermi and \hess binned event data sets as introduced in the previous sections.
As the centroid of the \gam-ray emission is slightly offset from the pulsar position (cf.\ Sect.~\ref{sec:ext_measurement}), we shifted the model prediction spatially such that its centre coincides with the centroid of the \gam-ray emission in order to ensure a proper convergence of the fit.
This does not affect the predicted spectrum in a noticeable way.
In the combined fit, all model components except the Crab Nebula SSC model (i.e.\ the background models for \hess and the Galactic and isotropic diffuse emission models as well as the remaining sources in the ROI for \fermi) are frozen to pre-fitted values.
Additionally, we fixed some of the model parameters to values employed by \citet{Dirson2023}, specifically all parameters relating to the dust model and the shock radius.
The fixed parameters are marked as such in Table~\ref{table:parameters}.

As the synchrotron photon field is a seed field for the IC up-scattering, we also need to ensure consistency with the observed synchrotron extensions which are measured with higher precision compared to the IC extensions.
Therefore, we also included archival flux and extension measurements from radio frequencies to X-ray energies, taken from \citet{Dirson2023}\footnote{The data points are available at \url{https://github.com/dieterhorns/crab_pheno}, where we have omitted points that also do not appear in Figs.~1 and~7 of \citet{Dirson2023}.}, to constrain the synchrotron flux predicted by our models (see also Fig.~\ref{fig:sed_intro}).
This is done by computing a $\chi^2$-value between the predicted and observed synchrotron flux and extensions and adding this value as a penalty term to the total $\mathcal{C}$-value of the fit of the IC spectrum,
\begin{linenomath*}
\begin{equation}
\label{eq:lik-sum}
    \mathcal{C}_\mathrm{tot} = -2 \ln\mathcal{L}_\mathrm{tot} = \mathcal{C}_\mathrm{IC} + \chi^2_\mathrm{SYN,flux} + \chi^2_\mathrm{SYN,ext}\,,
\end{equation}
\end{linenomath*}
where $\mathcal{C}_\mathrm{IC}=-2\ln\mathcal{L}_\mathrm{IC}$ is the Cash statistic of the model fitted to the \fermi and \hess data.
Lacking a detailed understanding of the systematic uncertainties associated with each of the synchrotron flux measurements, we added in quadrature to the uncertainty specified for each data point a relative systematic uncertainty of 7\% \citep[][see also Sect.~\ref{sec:systematics}]{Meyer2010}.

Compared to the analysis in \citet{Dirson2023}, we used data from only two instruments to constrain the IC emission.
However, both the \fermi and \hess data sets analysed here are much larger compared to those utilised by \citet{Dirson2023}.
Furthermore, we are able to include the extension of the IC nebula self-consistently in the fit because we use the three-dimensional information of the \fermi and \hess event data (i.e.\ a mis-match in extension would lead to strong residuals in the fit).
In order to achieve the largest possible overlap of the energy ranges covered by the \fermi and \hess data, we included the \hess mono data set in the combined analysis.
This is acceptable despite the systematic uncertainties associated with the PSF of the mono data set because this data set contributes to the fit mostly in a relatively narrow energy range ($\sim$\SIrange{240}{560}{GeV}), and we do not attempt to measure the extension as a function of energy here (in contrast to the analysis presented in Sect.~\ref{sec:ext_measurement}).

\begin{figure*}
    \sidecaption
    \includegraphics[width=12cm]{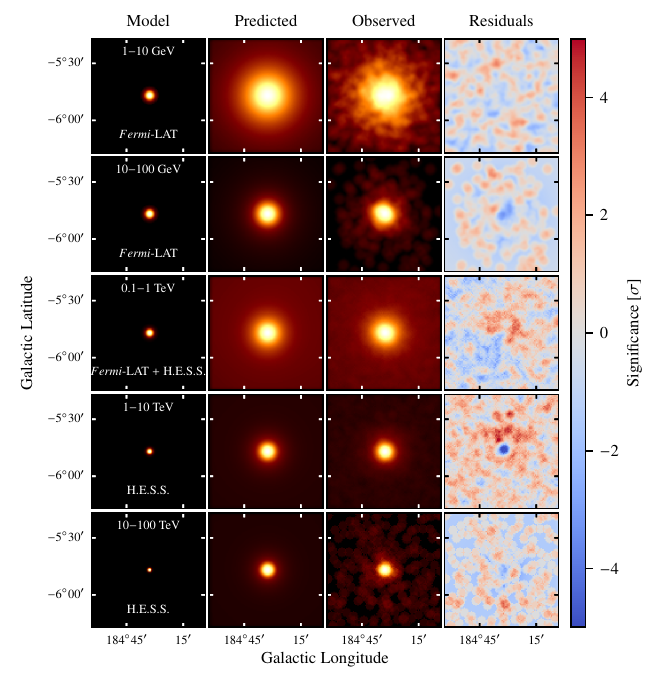}
    \caption{
        Overview showing the best-fit \vbm and the observed data in the IC regime, in five different energy bands.
        Each panel displays a $1^\circ\times 1^\circ$ cutout around the position of the Crab Nebula.
        The first column gives the best-fit model prior to folding with the IRFs.
        The second column illustrates the predicted number of events based on source and background models (folded with the IRFs), smoothed with a top-hat kernel of $2.4'$ radius.
        The third column shows the observed number of events with the same smoothing.
        Finally, the fourth column displays residual significance maps (see main text for details).
    }
    \label{fig:residuals}
\end{figure*}

Figure~\ref{fig:residuals} shows the best-fit spatial models and residuals in five different energy bands for the \vbm.
The residual maps show the deviations between the total (i.e.\ background + source) model prediction and the observed number of events in terms of $\sqrt{\mathrm{TS}}$, where $\mathrm{TS}\equiv -2\ln(\mathcal{L}/\mathcal{L}_\mathrm{max})$ and $\mathcal{L}_\mathrm{max}$ is the likelihood corresponding to a model that perfectly predicts the observed number of events in each pixel.
While not perfectly compatible with purely statistical fluctuations, the residual maps indicate that we achieve a satisfactory description of the extension of the nebula in the part of the IC spectrum probed with \fermi.
However, in the part probed with \hess, especially in the \SIrange{1}{10}{TeV} range, we observe stronger residuals.
The extension of the Crab Nebula is strongly constrained by the synchrotron measurements and appears too small when compared with the \hess data, leading to negative residuals at the position of the Crab Nebula that are surrounded by an unaccounted-for excess.
For the other two models this trend is similar, with even slightly larger residuals.
This indicates that our models cannot simultaneously describe both the synchrotron and IC extensions together with the broadband SED (however, see Sects.~\ref{sec:ext} and~\ref{sec:systematics} for a discussion of systematic uncertainties).

\subsection{Discussion of the models}
\label{sec:model_discussion}

\begin{table*}
    \caption{
        Overview of $\mathcal{C}$ and $\chi^2$ values obtained in the fits of our physical models.
    }
    \label{table:cstats}
    \centering
    \begin{tabular}{l c c c c c c}
        \hline\hline
        Model & $\Delta \mathcal{C}_\mathrm{tot}$ & $\Delta \mathcal{C}_\mathrm{IC}$ & $\chi^2_\mathrm{SYN,flux}$ (\#) & $\chi^2_\mathrm{SYN,ext}$ (\#) & $N_\mathrm{par}$ & $\Delta$AIC\\
        \hline
        \vbm & 0 & 0 & 152.7 (194) & 60.3 (14) & 16 & 0\\
        \cbm & 374.0  & 264.1 & 246.0 (194) & 76.9 (14) & 15 &  372.0\\
        Kennel \& Coroniti & 292.6 & 143.0  & 285.5 (194) & 77.1 (14) & 12 & 284.6\\
        \hline
    \end{tabular}
    \tablefoot{
        The $\Delta \mathcal{C}_\mathrm{tot}$, $\Delta \mathcal{C}_\mathrm{IC}$ (cf.\ Eq.~\ref{eq:lik-sum}), and $\Delta$AIC values are given with respect to the \vbm model.
        The $\chi^2$-values are given together with the number of flux and extension points included in the fit.
        $N_\mathrm{par}$ denotes the number of free parameters of the model.
        See main text for a definition of the different quantities.
    }
\end{table*}

We now turn to the discussion of the physical models and how well they fit our data.  
For each of the models, we give a quantitative overview of the fit results in Table~\ref{table:cstats}, that is, we provide the Cash statistic minimised by the fit as well as the number of fitted parameters and the number of synchrotron data points.
An overview of all best-fit model parameters is provided in Table~\ref{table:parameters}.
The complexity of the models prevents a detailed assessment of the uncertainties on the model parameters.
We therefore note that the best-fit values should not be regarded as measurements of the corresponding physical quantities, but simply as the set of values that yields the best agreement with the data.

Of the three models tested in this work, we find that the \vbm provides the best fit to the data.
A comparison between the \cbm and the \vbm shows that the former is strongly disfavoured with respect to the latter, both in the synchrotron and the IC part of the fit.
Since the models are nested (i.e.\ the \vbm can be reduced to the \cbm for a particular choice of parameter values), we can perform a likelihood ratio test and invoke Wilks' theorem \citep{Wilks1938} to infer that the \vbm is preferred by $\sqrt{\Delta\mathcal{C}_\mathrm{tot}}\approx 19.3$ standard deviations.
On the other hand, a comparison of the K\&C model and the \vbm by means of a likelihood ratio test is not possible, since these models are not nested.
Instead, for a statistical comparison of the K\&C model and the \vbm, we use again the Akaike information criterion, now defined as $\mathrm{AIC}=\mathcal{C}_\mathrm{tot}+2N_\mathrm{par}$.
We conclude that the \vbm is also strongly favoured over the K\&C model, with a difference in AIC of~$284.6$.

As a caveat to these comparisons, we note that the dust model has been optimised for the \vbm only, as a result of adapting the dust parameters from \citet{Dirson2023}.
While the other models would potentially benefit from an additional optimisation of the dust parameters, we expect this effect to be minor and to have no impact on the conclusions drawn in this study.
In the following, we discuss the models in terms of their predictions for the magnetic field, the broadband SED, and the nebula extension.

\subsubsection{Magnetic field}
\label{sec:b-field}

As already stated in \citet{Dirson2023}, the VHE data provide the strongest constraint on the shape of the magnetic field.
For the \vbm, in which the $B$-field is parameterised as $B(r)=B_0(r/r_s)^{-\alpha}$, we find best-fit values of $\alpha=0.47$ and $B_0=\SI{256}{\micro\gauss}$.
The steepness of the radial profile of the magnetic field is thus compatible with the value of $\alpha=0.51\pm0.03$ found by \citet{Dirson2023}, who have used other VHE data sets in their fit.

With the K\&C model we find a magnetisation of $\sigma=0.0214$, which is about four to seven times larger compared to previous estimates \citep{Kennel1984,Meyer2010}.
Such high values of $\sigma$ would lead to a high flow speed of $\gtrsim$ \SI{5000}{\kilo\meter\per\second} at the interface between the nebula and the supernova remnant and a predicted shock radius of $r_s \approx \sqrt{\sigma} r_0 \approx \SI{0.3}{pc}$, where $r_0 \approx \SI{2}{pc}$ is the size of the nebula \citep{Kennel1984}.
This is at odds with observed expansion and flow speeds of the order of $\sim$\SI{2000}{\kilo\meter\per\second} and the shock radius at $r_s = 0.13\,\mathrm{pc}$.
A value of $\sigma$ compatible with these measurements is ruled out by our analysis with high significance, as it would lead to a much larger value of $\mathcal{C}_\mathrm{tot}$ and thus be disfavoured by the AIC (cf.~Sect.~\ref{sec:model_discussion}).

\begin{figure}
    \centering
	\includegraphics[width=0.99\linewidth]{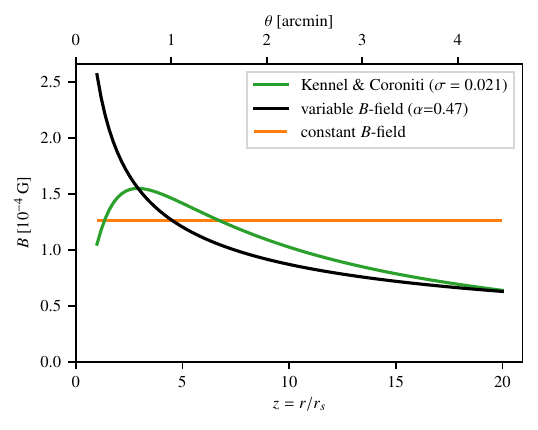}
	\caption{
         Magnetic field strength as predicted by the different models, as function of the distance to the nebula centre in units of the shock radius $r_s$.
         For the \vbm $B\propto r^{-\alpha}$, with $\alpha$ given in the legend.
         For the K\&C model the magnetisation $\sigma$ is specified in the legend.
	}
	\label{fig:b-fields}
\end{figure}

Figure~\ref{fig:b-fields} shows that the $B$-field profiles of the K\&C model and the \vbm agree quite well in the outer regions of the nebula ($r>5r_s$).
At the shock radius, the \vbm predicts more than twice the magnetic field strength compared to the K\&C model.
We can also compute a magnetisation for the \vbm by comparing the energy in the $B$-field against the energy in particles as a function of distance to the centre.
We find a value of $\sigma\approx0.1$ at the shock radius and $\approx0.03$ averaged across the whole nebula.

As discussed before, the \cbm performs significantly worse, implying that a $B$ field that decreases with distance from the centre is clearly preferred.
This confirms the findings of \citet{Dirson2023}, who have reached a similar conclusion.

\subsubsection{Broadband SED}
\label{sec:sed}

\begin{figure*}
    \includegraphics{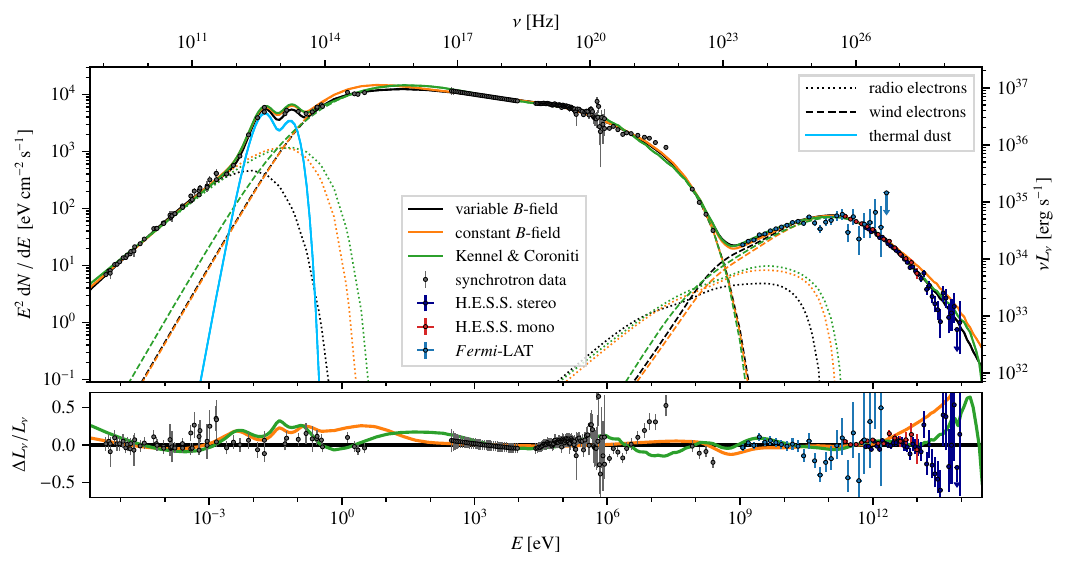}
    \caption{
        SED of the Crab Nebula, together with the predictions of the three models.
        The dotted lines show the emission produced by the radio electrons, while dashed lines are for the wind electrons (where synchrotron emission from all electrons is always included as IC seed photon field).
        The total emission is shown by the solid lines.
        The light blue line denotes infrared dust emission (identical for all models).
        The synchrotron data points are the same as in Fig.~\ref{fig:sed_intro}, with a 7\% systematic uncertainty added in quadrature.
        The IC data are from this work (cf. Fig.~\ref{fig:hess_sed}~\&~\ref{fig:fermi_sed}).
        The bottom panel shows the ratio of the data points as well as the constant $B$-field and K\&C model with respect to the \vbm.
    }
    \label{fig:sed_combined}
\end{figure*}

We compare the synchrotron flux prediction of the models in Fig.~\ref{fig:sed_combined}.
Even though the parametrisation of the radio electrons is identical for all models, minor differences can be seen in the radio part of the SED ($<$\SI{3}{\tera\hertz}).
Compared to the \vbm, both the K\&C model and the \cbm prefer a different transition between radio and wind electrons, that is, a larger cutoff energy for the radio electrons.
This leads to a slight mis-match in the SED, but is required to fit the spatial extension in the IR and optical regime (see below).
Since the parametrisation of the transition is not physically motivated, we will draw no further conclusions from the differences between the models in that regime.

The X-ray to \gam-ray part of the synchrotron spectrum is not described well by any of the models.
Specifically, the data between \SIrange{e20}{e22}{Hz} suggest a slightly harder spectrum followed by a stronger cutoff, which is not matched by the models.
Possible reasons for this could be the parametrisation of the super-exponential cutoff at $\gamma_\mathrm{w,max}$, or constraints resulting from the measured extension in this part of the spectrum, but also in the high-energy part of the IC spectrum.
Comparing the models among each other, we observe a very good agreement between the static models, while the K\&C model behaves slightly differently.
This is most likely due to the differences in the parametrisation of the wind electron distribution.
The first break in the spectrum, just beyond $\gamma_\mathrm{w,min}$, helps the static models with the transition between the UV and X-ray flux measurements.
The K\&C model, on the other hand, has only one break in the wind spectrum as the attempt of adding a second break resulted in no significant improvement.
Instead, the wind electrons in the K\&C model follow the form $(\gamma_\mathrm{w,min}+\gamma)^s$ (compared to $\gamma^s$ for the static models), which has a similar effect.
It should be noted that the K\&C model requires a harder electron index after the break to achieve agreement with the hard X-ray spectrum. 
Specifically, the spectral index of the electron spectrum hardens from 2.87 to 2.32, which is difficult to explain in the framework of the MHD flow model.

\begin{figure*}
    \centering
	\includegraphics{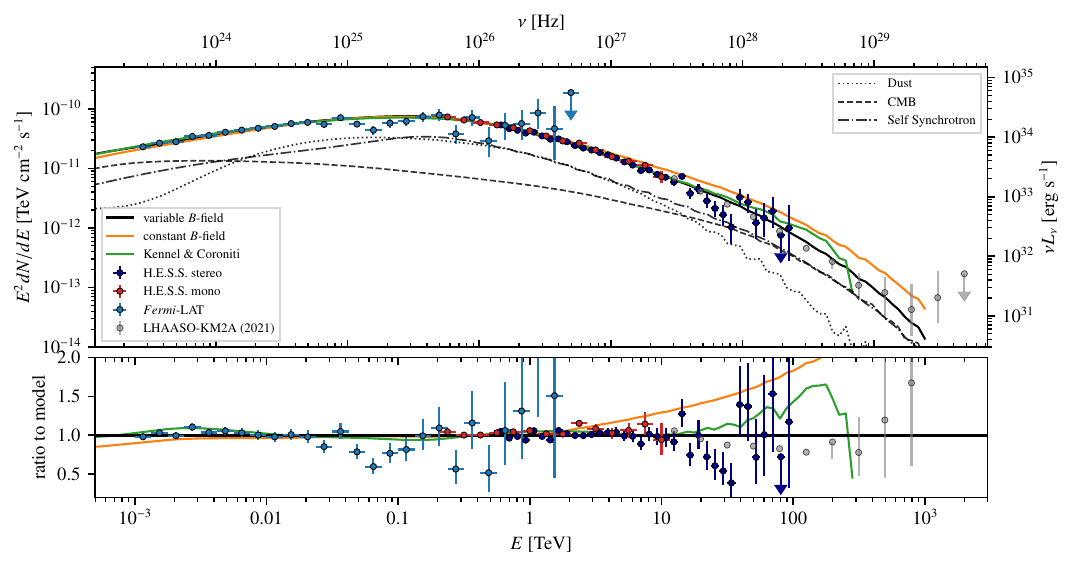}
	\caption{
        IC flux predictions of the three models, together with the \fermi and \hess flux points derived in this work.
        The LHAASO flux points, taken from \citet{LHAASO2021}, are only shown for comparison.
        The dashed, dotted and dash-dotted lines show the individual contributions of seed photon fields for the \vbm.
	    The bottom panel shows the ratio of the data points as well as the constant $B$-field and K\&C model with respect to the \vbm.}
	\label{fig:ic-spec}
\end{figure*}

A zoom-in of the IC part of the SED is shown in Fig.~\ref{fig:ic-spec}.\footnote{We note that the unsteady behaviour of the K\&C model at the highest energies is due to numerical instabilities and therefore not physical.}
We stress again that the models have not been fitted to the \fermi and \hess flux points shown in this figure (and Fig.~\ref{fig:sed_combined}), but to the binned event data of these instruments directly; the flux points are shown for comparison only here.
As indicated by the error bars of the flux points, the data provide the strongest constraints at energies of a few GeV, and at around \SI{1}{TeV}.
As a result, the model predictions are very similar up to an energy of a few~TeV.
The main differences appear at energies above \SI{10}{TeV}, where in particular the \cbm over-predicts the measured flux.
The electrons responsible for this flux are also producing the highest-energy synchrotron flux, where we find an excellent agreement between the static models.
However, the \vbm predicts the same synchrotron flux with a softer electron spectrum and lower high-energy cutoff (cf.\ Fig.~\ref{fig:el-spec}), because of the higher $B$-field strength near the shock radius.
This leads to a stronger reduction of the IC flux with respect to the synchrotron flux.
When extrapolating the IC flux above 100\,TeV, the \vbm agrees with the LHAASO data \citep{LHAASO2021} up to $\sim$\SI{1}{PeV}, even though these data are not included in the fit.

\begin{figure}
    \centering
	\includegraphics[width=0.99\linewidth]{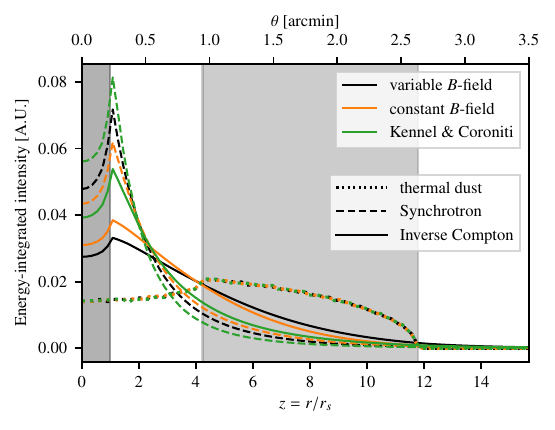}
	\caption{
        Energy-integrated intensity predicted by the three models, as a function of distance from the centre of the Crab Nebula.
        The solid, dashed, and dotted line correspond to IC, synchrotron, and dust emission, respectively; the colours refer to the three different tested models.
        The intensity has been integrated above \SI{1}{meV} for the synchrotron emission and above \SI{1}{GeV} for the IC emission.
        The dark grey band up to $z=1$ indicates the region up to the shock radius, in which no emission is predicted by the models.
        The light grey band shows the dust shell, in which all the dust emission happens.
        The profiles show non-zero emission inside the shock region and dust emission outside the dust shell because the emissivity is integrated along the line of sight.
    }
	\label{fig:profiles}
\end{figure}

The seed photon field contributions are shown in  Fig.~\ref{fig:ic-spec} for the \vbm, but are very similar for the other models.
It can be seen that the SSC component follows the spectral shape across all energies.
The dust component is similar in strength to the SSC component between $\sim$\SI{10}{GeV} and $\sim$\SI{10}{TeV}, while being suppressed at lower and higher energies.
The CMB component, on the other hand, behaves in exactly the opposite way.
This rather complex behaviour emerges as the seed photon fields do not only differ in their spectrum, but also their morphology.
While the CMB is constant at all radii, the SSC component approximately follows a Gaussian profile and the dust component is modelled as a shell around the nebula.
As an example, at several tens of TeV the extension of the electron population becomes smaller than the inner radius of the dust shell, leading to a reduced contribution to the IC emission from dust photons at these energies.
Another way of illustrating the spatial distribution of the emission in close vicinity to the centre is displayed in Fig.~\ref{fig:profiles}, where we show the energy-integrated intensity profiles (emissivity integrated along the line of sight) for the thermal dust, synchrotron, and IC emission, as predicted by our best-fit models.
It is evident that, compared to the \cbm, the two models with a variable $B$-field predict a narrower profile for the synchrotron emission.
Despite this, the \vbm exhibits the flattest IC profile of all models.
In other words, the rapidly decreasing $B$-field allows for a more extended electron distribution while maintaining a narrow synchrotron profile.
This leads to a larger predicted extension in the IC regime, which matches the observations best.

\subsubsection{Nebula extension}
\label{sec:ext}
\begin{figure}
    \centering
	\includegraphics[width=0.99\linewidth]{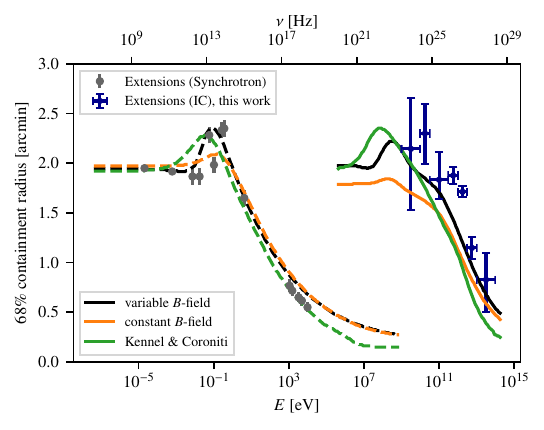}
	\caption{
        Comparison of the 68\% flux containment predicted by the models as function of energy (dashed for the synchrotron part and solid for the IC part).
        The grey synchrotron extension points are taken from \citet{Dirson2023} and references therein.
        The error bars of the data points in the IC domain denote statistical uncertainties only.
	}
	\label{fig:extensions}
\end{figure}

\begin{figure}
    \centering
	\includegraphics[width=0.99\linewidth]{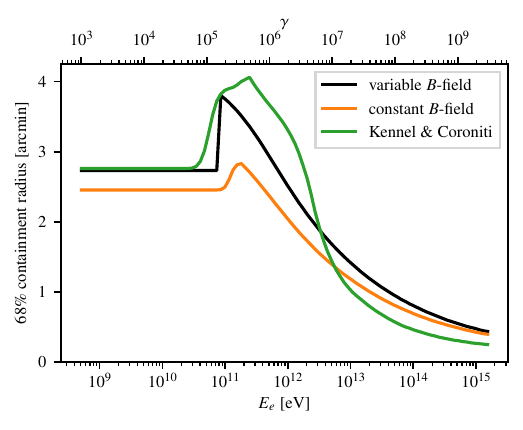}
	\caption{
        Comparison of the 68\% containment radii of the electron energy densities of the three models. 
        The radio electrons dominate the low-energy part of the spectrum where the extension is not energy-dependent.
        The extension of the wind electrons changes with energy, following, in case of the static models, a simple power law.
        The spectral transition of the two distributions is shown in Fig.~\ref{fig:el-spec}.
	}
	\label{fig:electron-extensions}
\end{figure}

In Fig.~\ref{fig:extensions}, we show the extension of the nebula predicted by the models from the radio to the \gam-ray domain, and compare this to the corresponding measurements.
The extensions of the respective electron distributions are shown in Fig.~\ref{fig:electron-extensions}.
At energies for which the radio electrons dominate ($\lesssim$10\,meV for the synchrotron and $\lesssim$10\,MeV for the IC part), the morphology of the model is independent of energy.
When the more extended low-energy wind electrons start to dominate, the maximal extension is reached in the optical (around 100\,MeV for the IC part).
That the low-energy wind electrons exhibit a larger extension than the radio electrons can be physically motivated by energy-dependent diffusion, see for example \citet{Tang2012}.

First, we note that our measurement of a decreasing extension with increasing energy in the IC regime is generally well in line with the model predictions, and thus not a surprising result.
Ultimately, this is a consequence of the already observed decrease of the size of the nebula in the synchrotron domain.
Comparing the models in detail, we find that only the K\&C model fully predicts the strong decrease in extension towards the X-ray energies, whereas the static models seem to be influenced more by the larger extensions measured at \gam-ray energies.
Accordingly, the K\&C model under-predicts the extension across most of the IC regime, while the \vbm\ -- albeit still predicting a too-small extension -- provides a better match to the data here.
We furthermore find strong correlations between the spectral and spatial parameters, as a narrower spatial distribution also decreases the total number of electrons and thus the \gam-ray flux at the corresponding energies.
It does not seem possible for any of the models -- for the given parameters of the electron spectrum and the seed photon fields -- to predict the measured IC extensions together with the broadband SED.
We see two possible reasons for that.

First, the discrepancies could be due to simplifying assumptions made by the models.
For instance, all models assume spherical symmetry, which is clearly an oversimplification. 
Additionally, three-dimensional and non-ideal MHD calculations suggest deviations from the predictions of the K\&C model \citep{Porth2013,Bucciantini2017,Tanaka2018,Lyutikov2019,Luo2020}.
In particular, non-ideal MHD models which predict turbulent magnetic fields to be present can resolve a number of problems of the K\&C model \citep{Luo2020}.
This would necessarily alter the magnetic field structure in the nebula and thereby the relation between synchrotron and IC photons.
This could potentially lead to a better agreement with the data.
However, a spatially resolved calculation of such models is so far not available and the current non-ideal MHD models cannot reproduce the full SED in all its details.
Another option would be the presence of an additional electron population that extends beyond the synchrotron nebula, for example electrons that have already escaped the nebula or electrons moving inwards from the so-far unobserved outer shock. 
The IC scattering of such a population with the CMB could potentially lead to a larger IC extension. 

The second possibility is systematic uncertainties, for example a mis-modelling of the \hess PSF beyond the estimated systematic uncertainty, which could cause the extension measured in the IC regime to appear larger than it actually is.
Systematic uncertainties could also include an unaccounted factor between true and reconstructed energies of the IC photons.
However, the required magnitude of such systematic effects appears to be unrealistically high to bring the measurements in agreement with the model predictions. 
We elaborate on this in the next section. 

\section{Systematic uncertainties and energy scale}
\label{sec:systematics}
Several systematic uncertainties could impact our analysis. 
For example, the absolute energy scales of \fermi and \hess could differ, as the energy reconstruction of the two instruments is calibrated in different ways.
Specifically, for \fermi a beam test in combination with a measurement of the geo-magnetic cutoff was used to infer the absolute energy scale with an accuracy of $^{+2}_{-5}$\% \citep{FermiLAT_Escale_2012}. 
Reaching this level of accuracy is not possible with \hess, where the atmosphere effectively is part of the detector and variations in atmospheric conditions can lead to a change in the number of Cherenkov photons reaching the telescopes.
While in our selection of observation runs we have excluded observations carried out under sub-optimal conditions, distributions of the `Cherenkov Transparency Coefficient' \citep{Hahn2014} for the selected observations indicate that variations of about 10\% remain.
A difference in energy scale between the instruments could thus lead to a systematic shift between the spectrum of the Crab Nebula measured with \fermi and \hess \citep[cf.\ also][]{Nigro2019}.

\begin{table*}
    \caption{Fit quality with and without including systematic uncertainties.}
    \label{table:sys}
    \centering
    \begin{tabular}{l c c c}
        \hline\hline
         & w/o systematics & & w/ systematics\\
        Data set & $\chi^2$ (sys.\ error) & $N_\mathrm{dof}$ &  $\chi^2$ (sys.\ error)\\
        \hline
        \hess stereo (flux points) & 45.5 (none) & 35 & 33.6 (3\%) \\
        \hess mono (flux points) & 36.7 (none) & 13 & 12.1 (4\%) \\
        \fermi (flux points) & 41.7 (none) & 26 & 27.7 (10\%) \\
        Synchrotron (flux points) & 235.0 (5\%) & 194 & 152.7 (7\%) \\
        \hline
        \fermi + \hess (extension) & 121.5 (none) & 6 & 8.9 (see text) \\
        Synchrotron (extension) & 60.3 (n.a.) & 13 & 13.0 (8$''$) \\
        \hline
    \end{tabular}
    \tablefoot{
        The $\chi^2$ values are calculated for the respective data set and the \vbm.
        For each data set, we conservatively estimate the number of degrees of freedom ($N_\mathrm{dof}$) as the number of data points minus one.
        The magnitude of the systematic uncertainties (specified in parentheses) are chosen such that a good fit quality is obtained, that is, $\chi^2 \sim N_\mathrm{dof}$, and thus do not represent estimates of the actual systematic uncertainty of each data set.
        The first section of the table summarises the flux measurements, where the systematic uncertainty is specified as a relative error on the flux level.
        The extension uncertainty, given in the second section of the table, is increased by a constant error for each data set (see main text for details).
        We note that we still add a systematic uncertainty of 5\% to the synchrotron flux points for the `w/o systematics` case, as the $\chi^2$-value would otherwise increase to $>10^6$, due to the small statistical uncertainties of some of the points.
        Similarly, the extension measurements in the synchrotron regime already contain systematic uncertainties, which are however not easily separable and thus not stated explicitly here (see \citealt{Dirson2023} for details).
    }
\end{table*}

One method of incorporating a scale factor $\eta$ between the energy scales is to evaluate the \hess source model at a modified energy $\eta\cdot E$, while the evaluation of the \fermi source model is unchanged \citep{Meyer2010}.
Leaving $\eta$ as an additional free `nuisance' parameter during the fit determines the difference in energy scale between the two instruments.
This method, however, depends on the assumed spectral model, and a model that fits the data poorly can falsely indicate a scale factor significantly different from unity if this improves the agreement with the data.
For example, adopting a log-parabola spectrum leads to a factor $\eta = 0.79\pm 0.03$ simply because the model is not suitable for the whole energy range.
Other models with the possibility of a sharper break in the spectrum result in values $\eta > 1$.
For the \vbm, we find $\eta=1.038$, with only a very minor improvement of the fit ($\Delta \mathcal{C}_\mathrm{tot}\approx 2$, implying that $\eta$ is consistent with a value of 1).
As we cannot be certain about the correctness of our models, we therefore chose the approach of fixing $\eta=1$ in our final analysis.
This is further supported by looking at the integrated flux in the overlapping energy range between the \fermi data and the \hess mono data set (i.e.\ \SIrange{0.24}{1.80}{TeV}).
In this energy range, we do not see an indication for the energy shift as the integrated fluxes agree within statistical uncertainties.
Specifically, summing the flux points in this range we find a \fermi flux of \SI{1.9(3)e-10}{\per\square\centi\meter\per\second}, which is consistent with the \hess mono flux of \SI{2.09(2)e-10}{\per\square\centi\meter\per\second} (statistical uncertainties only).

Nonetheless, we have aimed to estimate the systematic uncertainty associated with our flux measurements.
To do so, we assessed the quality of the fit of the \vbm by means of a $\chi^2$ test, and determined the magnitude of a potential systematic error that would lead to an acceptable fit quality.
For the spectral measurements in the \gam-ray regime, we added a constant percentage of the flux in quadrature to the statistical uncertainty.
To calculate the $\chi^2$ values for the \gam-ray data sets, we used the asymmetric errors of each flux point, meaning that we used the positive error if a point falls below the model prediction and vice versa.
In the synchrotron range, the systematic error is also added as a percentage of the measured flux in quadrature to the statistical errors (as in the fitting process, see Sect.~\ref{sec:modelling}).
For the synchrotron extension measurements, the assumed systematic error is a constant angle.
Finally, in the IC regime, we used the systematic errors of the extensions previously cited in the single-instrument analyses.
That is, at energies for which the \fermi data dominate we added an error of $^{+0.54}_{-0.48}\,'$ (cf.\ Sect.~\ref{sec:fermi_data_analysis}) and at energies for which the \hess data dominate we added an error of $^{+0.21}_{-0.24}\,'$ (cf.\ Sect.~\ref{sec:hess_data_analysis}).
Again, all of the systematic uncertainties have been added in quadrature to the statistical ones.

Table~\ref{table:sys} summarises the $\chi^2$ values for the \vbm with and without the added systematic uncertainties, together with the estimated number of degrees of freedom ($N_\mathrm{dof}$). 
For the case of a single data set to which a model has been fitted, $N_\mathrm{dof}$ is equal to the number of data points minus the number of model parameters (when neglecting correlations between the parameters).
We note that in the present case, with the models being fitted to multiple data sets at once, it is not straightforward to properly state the exact $N_\mathrm{dof}$ for each of the individual sets.
The reason for this is that the models have not been adjusted to any of the individual sets alone, but to their combination.
The high degree of correlation between some of the model parameters presents an additional complication.
In Table~\ref{table:sys}, we therefore provide as a very conservative estimate of $N_\mathrm{dof}$ the number of points in the data set minus one (taking into account one overall scale factor that remains even if all model parameters are perfectly correlated).
We find that without added systematic uncertainties, the $\chi^2$ values are all well above $N_\mathrm{dof}$, indicating an unsatisfactory model description or the presence of systematic uncertainties.
For the \hess stereo and mono flux points, an additional systematic flux uncertainty of 3\% and 4\%, respectively, leads to an acceptable fit quality.
This flux uncertainty would correspond to an uncertainty on the energy scale of the instrument of 1.2\% and 1.6\%, respectively.
For the \fermi data set, a systematic flux uncertainty of 10\% is required, corresponding to a 5\% uncertainty on the energy scale.
We stress that these values are not generally valid estimates for the energy scale uncertainty of \hess and \fermi, but only the level of systematic scaling required to achieve an acceptable goodness-of-fit in our analysis.
The variations are, however, well within the typically adopted systematic uncertainties of the two instruments\footnote{See e.g.\ \citet{HESS2006} for \hess and \url{https://fermi.gsfc.nasa.gov/ssc/data/analysis/LAT_caveats.html} for \fermi.}.
The $\chi^2$ value of the synchrotron flux points depends sensitively on the assumed systematic error because of the relatively small statistical uncertainties on some of the points.
We find that adding a 7\% flux uncertainty leads to a $\chi^2$ value well below $N_\mathrm{dof}$, while 5\% is not sufficient.
Compatible best-fit parameters are obtained with either assumed systematic error.

Regarding the extension of the nebula, it is evident that without consideration of systematic uncertainties, the predicted extension in both the IC and synchrotron regime is strongly at odds with the measurements.
For the \fermi and \hess data sets, taking into account the previously estimated systematic uncertainties leads to a marginally acceptable fit quality.
The assumed error bars of the synchroton data points, on the other hand, already contain systematic uncertainties \citep[see][]{Dirson2023}.
Nevertheless, an additional uncertainty of 8$''$, which is approximately twice as large as the uncertainty used in the fit ($\sim$4$''$ on average), would be required to achieve $\chi^2\approx N_\mathrm{dof}$.
This illustrates again the previously discussed inability of the model to simultaneously describe the SED and the extension measurements.
The simplifying assumptions of the model already discussed in the preceding Sect.~\ref{sec:ext}, as for example radial symmetry and a simple power law describing the electron distribution size evolution with energy, may not be justified.

\section{Conclusion}
\label{sec:conclusion}
In this work, we present a joint \fermi and \hess analysis of the Crab Nebula, carried out with the open-source analysis package \gp.
Below, we first summarise our findings regarding the technical part of our work -- related to the joint fit of the \fermi and \hess data -- before we discuss the implications of our analysis on emission models of the Crab Nebula.

First, we have demonstrated that applying the open-source analysis package \gp to \fermi and \hess data, we are able to obtain results that are consistent with the traditionally used analysis chains.
Specifically, for \fermi, we show that we can almost exactly reproduce results obtained with the \fermipy package after some minor adjustments in the data reduction procedure.
In the context of this work, \gp has the advantage that it enables a combination of the \fermi data with that of \hess, and it offers the possibility to include customised physical models in the analysis.
For \hess, on the other hand, \gp allows us to apply the three-dimensional likelihood analysis method, that is, to model the spectrum and morphology of the Crab Nebula simultaneously under full consideration of the Poisson event statistics.
A key ingredient for this is a three-dimensional model for the residual hadronic background.
We find consistent results with more traditional methods that estimate this residual background from the analysed observations themselves.
Second, by combining the \fermi and \hess data in a joint-likelihood analysis, we provide a fully consistent spectro-morphological analysis of the Crab Nebula over five decades in energy.
We confirm previously found indications for an extension of the nebula in the IC regime that decreases with increasing energy -- a result that agrees well with theoretical expectations.
We discuss the possibility of taking into account possible systematic effects (e.g.\ different energy scales between instruments) in the form of nuisance parameters in the likelihood fit.
Finally, to be able to fit physical emission models to our data that are consistent with other MWL observations, we demonstrate that it is possible to take into account MWL constraints through the addition of penalty terms to the fit statistic.

This has enabled us to consistently model the synchrotron and IC part of the emission from the Crab Nebula.
Specifically, we have confronted three phenomenological models for the non-thermal emission of the Crab Nebula with our measurements: a \cbm based on \citet{Meyer2010}, a \vbm based on \citet{Dirson2023}, and the K\&C model by \citet{Kennel1984}.
We note that these models are open-source and publicly available\footnote{See \url{https://github.com/me-manu/crabmeyerpy} for the implementation we have used.}.
All models predict the \gam-ray emission not only as a function of energy, but also as a function of the distance from the centre of the nebula.
Hence, by projecting the symmetric model on the sky, we were able to fit a three-dimensional model of the Crab Nebula emission directly to the measured event data from \fermi and \hess.
We conclude that, within the range of the estimated systematic uncertainties, none of the models are able to describe the full SED together with the extension of the nebula at different energies.
We firmly rule out the \cbm, as it completely fails to describe the decreasing extension with energy and the spectrum at the same time.
The K\&C model describes the SED and the extension of the synchrotron nebula well, but under-predicts the size of the IC nebula.
Furthermore, it requires a spectral hardening in the electron spectrum to fit the X-ray data, and the best-fit value of the magnetisation $\sigma$ is at odds with the position of the shock and the expansion velocity of the nebula.
Finally, the \vbm achieves the best agreement with the data.
This indicates that the magnetic field strength in the nebula decreases with increasing distance from the pulsar.
However, compared to the model prediction, the new \hess measurements of the extension presented here are in tension with the extension measurements in the X-ray domain.
As a consequence, the \vbm does not achieve the same fit quality as in \citet{Dirson2023}.
In summary, even though we find strong evidence that the extension of the IC emission decreases with energy, it still appears too large compared to the distribution of the synchrotron photons, which are supposedly produced by the same electrons.
This could suggest more complicated spatial profiles of the electron distribution and the magnetic field as predicted, for example, in non-ideal MHD calculations~\citep[e.g.][]{Porth2013,Luo2020}.
Additionally, a non-spherical geometry (such as a toroidal or flattened profile) will also change the seed photon density, which could lead to a better agreement between the model predictions and data.

As a final note, we remark that the IC spectrum above $\sim$\SI{30}{TeV} is not well constrained by the data used in this work.
Hence, differences between the models occur predominantly in this regime.
In future, including data from instruments optimised for the highest energies, such as LHAASO, could help in placing further constraints on the models and in determining the shape of the magnetic field in more detail.
Moreover, data from the upcoming Cherenkov Telescope Array \citep[CTA;][]{CTA2018,Hofmann2023} with its unprecedented angular resolution will allow us to measure the extension of the Crab Nebula in the IC regime with higher precision.

\begin{acknowledgements}
We thank Matthew Kerr for providing the pulsar timing information for our \fermi data set.

The support of the Namibian authorities and of the University of Namibia in facilitating the construction and operation of \hess is gratefully acknowledged, as is the support by
the German Ministry for Education and Research (BMBF),
the Max Planck Society,
the German Research Foundation (DFG),
the Helmholtz Association,
the Alexander von Humboldt Foundation,
the French Ministry of Higher Education, Research and Innovation,
the Centre National de la Recherche Scientifique (CNRS/IN2P3 and CNRS/INSU),
the Commissariat \`a  l’\'energie atomique et aux \'energies alternatives (CEA),
the U.K. Science and Technology Facilities Council (STFC),
the Irish Research Council (IRC) and the Science Foundation Ireland (SFI),
the Knut and Alice Wallenberg Foundation,
the Polish Ministry of Education and Science, agreement no. 2021/WK/06,
the South African Department of Science and Technology and National Research Foundation,
the University of Namibia,
the National Commission on Research, Science \& Technology of Namibia (NCRST),
the Austrian Federal Ministry of Education, Science and Research
and the Austrian Science Fund (FWF),
the Australian Research Council (ARC),
the Japan Society for the Promotion of Science,
the University of Amsterdam and
the Science Committee of Armenia grant 21AG-1C085.
We appreciate the excellent work of the technical support staff in Berlin, Zeuthen, Heidelberg, Palaiseau, Paris, Saclay, T\"ubingen and in Namibia in the construction and operation of the equipment.
This work benefited from services provided by the \hess Virtual Organisation, supported by the national resource providers of the EGI Federation.

The \textit{Fermi}-LAT Collaboration acknowledges support for LAT development, operation and data analysis from NASA and DOE (United States), CEA/Irfu and IN2P3/CNRS (France), ASI and INFN (Italy), MEXT, KEK, and JAXA (Japan), and the K.A.~Wallenberg Foundation, the Swedish Research Council and the National Space Board (Sweden). Science analysis support in the operations phase from INAF (Italy) and CNES (France) is also gratefully acknowledged. This work performed in part under DOE Contract DE-AC02-76SF00515.

The authors gratefully acknowledge the compute resources and support provided by the Erlangen National High Performance Computing Center (NHR@FAU).
This research made use of the \textsc{Astropy} (\url{https://www.astropy.org}; \citealt{Astropy2013,Astropy2018,Astropy2022}), \textsc{matplotlib} (\url{https://matplotlib.org}; \citealt{Hunter2007}) and \textsc{iminuit} (\url{https://iminuit.readthedocs.io}; \citealt{Dembinski2020}) software packages.
\end{acknowledgements}

\bibliographystyle{aa}
\bibliography{references}

\begin{appendix}

\newpage
\section{Comparison with the reflected regions background method}
\label{appx-comp-hess}
\begin{figure}[hb]
    \centering
    \subfigure[]{
        \includegraphics[width=0.98\linewidth]{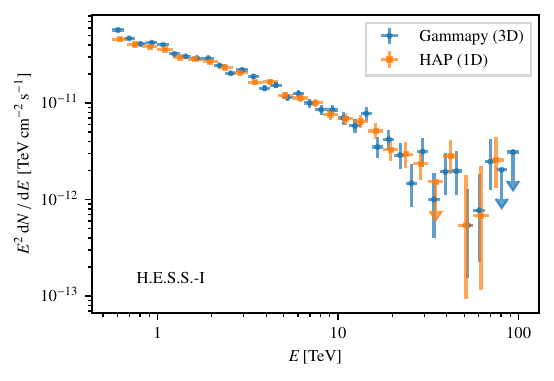}
    }
    \subfigure[]{
        \includegraphics[width=0.98\linewidth]{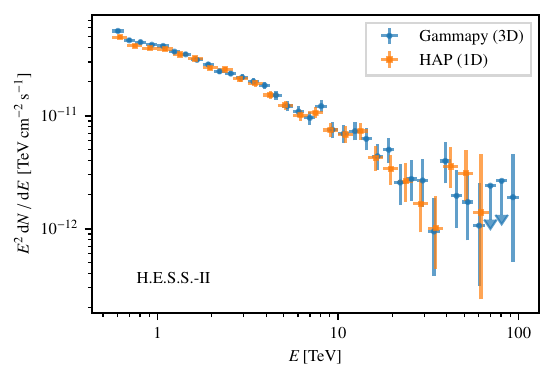}
    }
	\caption{
        Comparison of the Crab Nebula \gam-ray flux measured with \hess using different analysis techniques.
        The blue points with circle markers show the result as derived in the three-dimensional likelihood analysis with \gp described in this work.
        The orange points with square markers have been computed with the \hess analysis program (HAP), using a traditional one-dimensional analysis based on the `reflected background' method.
        (a) Data from \hess Phase-I (2004--2009, cf.\ Table~\ref{tab:hess_data_sets}).
        (b) Data from \hess Phase-II (2013--2015).
	}
	\label{fig:comp-hess}
\end{figure}

In this section we compare the \hess flux points derived with the three-dimensional (3D) likelihood analysis as described in this paper to those derived with a more traditional one-dimensional (1D) analysis, in which only the energy dimension is considered.
The main difference between the two methods lies in the estimation of the residual hadronic background.
While the 3D analysis requires a background model, the background for the 1D analysis is estimated using `Off' regions in the field of view \citep[`reflected regions method', see][]{Fomin1994,Berge2007}.
The 1D analysis has been carried out using the \hess-internal `HAP' pipeline.
Because the Crab Nebula is slightly extended we choose a comparatively large `On' region of $0.1^\circ$ radius.
For technical reasons, we also split the \hess stereo data set into two subsets for this test, corresponding to the different phases of the instrument.
The comparison of the derived spectra is shown in Fig.~\ref{fig:comp-hess}.
At low energies, the \gp points tend to be higher than those obtained with HAP.
We attribute this partly to the fact that the 1D analysis corrects for leakage of \gam rays outside of the On region under the assumption of a point-like source, whereas the Crab Nebula appears increasingly extended towards lower energies.

\clearpage

\section{Physical emission models for the Crab Nebula}
\label{appx-emission_model}
We model the broadband electromagnetic emission of the Crab Nebula through a radially symmetric steady-state model of the electron distribution, inspired by the constant $B$-field model introduced by \citet{Meyer2010}, which has been extended to a spatially \vbm in \citet{Dirson2023}.
An alternative description is the radially symmetric MHD flow solution obtained by \citet{Kennel1984}.
For comparison with the data, we need to calculate the predicted emission as a function of energy and angular separation between the line of sight and the nebular centre (this is equal to the two-dimensional sky position in the radially symmetric case).
The basic ingredients to do so are the radial profile of the magnetic field, $B(r)$, and the electron distribution, $n_\mathrm{el}(\gamma, r)\equiv dN_\mathrm{el}/d\gamma dV$, that is, the number of electrons per unit volume and Lorentz factor $\gamma$ as a function of the distance to the nebula centre.

Given $B(r)$ and $n_\mathrm{el}$, we calculate the spectral volume emissivity $j^\mathrm{sync}_\nu(\nu, r) \equiv dE / dt dV d\nu d\Omega$ for the synchrotron emission of a (pitch-angle averaged) chaotic magnetic field through \citep{Blumenthal1970,Aharonian2010}
\begin{linenomath*}
\begin{equation}
\label{eq:j-sync}
    j^\mathrm{sync}_\nu(\nu, r) = \frac{1}{4\pi}\frac{\sqrt{3} e^3 B(r)}{m_e c^2} \int\limits_1^\infty d\gamma n_\mathrm{el}(\gamma, r)~G\left(\frac{\nu}{\nu_c}\right)\,,
\end{equation}
\end{linenomath*}
where $\nu_c = 3 eB \gamma^2 / 4\pi m_e c$ is the critical frequency and $G(x)$ is given by 
\begin{linenomath*}
\begin{eqnarray}
    G(x) & = & \frac{x}{20}\left[\left(8 + 3x^2 \right)\left(K_{1/3}(x/2)\right)^2 + \right. \nonumber \\ & & \left.x K_{2/3}(x/2) \left(2K_{1/3}(x/2) -3 x K_{2/3}(x/2) \right)\right]\,,
\end{eqnarray}
\end{linenomath*}
with $K_\xi$ the modified Bessel function of kind $\xi$.
Throughout this paper, we use units of erg\,s$^{-1}$Hz$^{-1}$\,cm$^{-3}$\,sr$^{-1}$ for $j_\nu$.

For the dust emission, we follow \citet{Dirson2023}, who model the emission as a mixture of two dust populations at different temperatures $T_i$ and total masses $M_i$, $i=1,2$. 
The two populations are contained in a shell with inner and outer radii $r_\mathrm{dust,~in}$ and $r_\mathrm{dust,~out}$, respectively, with respect to the pulsar position (which is equal to the centre of the nebula in our model).
The density of dust within the shell is assumed to be constant. 
Assuming the two populations predominantly consist of amorphous carbon dust grains, the emissivity is given by
\begin{linenomath*}
\begin{equation}
\label{eq:j-dust}
    j^\mathrm{dust}_\nu(\nu, r) = \frac{3\kappa_\mathrm{abs}\big[\Theta(r - r_\mathrm{dust,~in}) -\Theta(r - r_\mathrm{dust,~out}) \big]}{4\pi (r_\mathrm{dust,~out}^3-r_\mathrm{dust,~in}^3)}\sum_{i = 1,2}M_i B_\nu(T_i)\,,
\end{equation}
\end{linenomath*}
where $\Theta(r)$ is the Heavyside step function, and $B_\nu(T)$ is the intensity of the black body emission at temperature $T$. 
The wavelength-dependent absorption coefficient is given by
\begin{linenomath*}
\begin{equation}
    \kappa_\mathrm{abs} = 2.15\times10^{-4}\,\mathrm{cm}^2\mathrm{g}^{-1}~\left(\frac{\lambda}{\mu\mathrm{m}}\right)^{-1.3}\,.
\end{equation}
\end{linenomath*}

Both the synchrotron and dust emission together with the CMB act as seed photon fields for IC scattering of the electrons producing the synchrotron emission. 
The seed photon density, $n_\mathrm{seed}^t$, $t=($sync, dust$)$, as a function of photon energy $\epsilon = h\nu$ is calculated through the integral \citep{Atoyan1996}
\begin{linenomath*}
\begin{equation}
    n_\mathrm{seed}^t (\epsilon, r) = \frac{r_0}{2hc}\frac{4\pi}{\epsilon}\int\limits_{r_s/r_0}^{1}dy \frac{y}{x}\ln\frac{x+y}{|x - y|} j_\nu^t(\nu, r_0 y)\,,\label{eq:nseed}
\end{equation}
\end{linenomath*}
in units of photons\,$\mathrm{eV}^{-1}\,\mathrm{cm}^{-3}$ and where $x = r/r_0$ and $y = r'/r_0$, with $r_0$ the radius of the nebula and $r_s = 0.13\,\mathrm{pc}$ \citep{Weisskopf2012} the shock radius until which no emission is assumed.\footnote{
One should note that the actual size of the nebula is unknown as the outer shock is not observed \citep[e.g.][]{Hester2008}.
The size of the observed synchrotron nebula is a free parameter in our fit (see below) and $r_0$ acts as an arbitrary upper integration bound that should be large enough to accommodate the different nebula sizes probed in the fit. Here, we set it to 3.6\,pc, which is twice the value assumed by \citet{Atoyan1996} for the visible nebula size.
}
The seed photon density for CMB photons is simply $n_\mathrm{seed}^\mathrm{CMB} = (4\pi / (hc)) B_
\nu(T_\mathrm{CMB}) / (h\nu)$.
We do not consider any additional photon fields such as optical line emission from the filaments \citep[e.g.][]{Meyer2010} or interstellar radiation fields, as these fields will give a subdominant contribution compared to the optical synchtrotron and infrared dust emission, respectively.

The IC emissivity for up-scattered photons at frequency $\nu$ is then calculated using the integral over the IC kernel $f_\mathrm{IC}$, which includes Klein-Nishina effects, and the seed photon density \citep{Blumenthal1970},
\begin{linenomath*}
\begin{eqnarray}
\label{eq:j-ic}
    j^\mathrm{IC}_\nu(\nu, r) & = & \frac{3\sigma_\mathrm{T}hc}{4}\frac{h\nu}{4\pi}\int\limits_1^\infty d\gamma\frac{n_\mathrm{el}(\gamma, r)}{\gamma^2} \nonumber \\
    & \times & \int\limits_0^\infty d\epsilon f_\mathrm{IC}(\nu, \epsilon, \gamma) \sum\limits_t \frac{n_\mathrm{seed}^t(\epsilon, r)}{\epsilon}\,,
\end{eqnarray}
\end{linenomath*}
with the usual expression for the full IC kernel,
\begin{linenomath*}
\begin{equation}
    f_\mathrm{IC}(\nu, \epsilon, \gamma) = 2q\ln q + (1 + 2q)(1-q) + \frac{1}{2}\frac{(\Gamma_\epsilon q)^2}{1 + \Gamma_\epsilon q}(1-q)\,,
\end{equation}
\end{linenomath*}
where $\Gamma_\epsilon = 4\epsilon\gamma/(mc^2)$ and $q = h\nu / (\Gamma_\epsilon(\gamma mc^2 - h\nu))$.
The Thompson limit corresponds to $\Gamma_\epsilon \ll 1$ and IC scattering only occurs for $1/(4\gamma^2) \leqslant q \leqslant 1$. 

The integrals are evaluated numerically and the code, fully written in \texttt{python}, is available online\footnote{\url{https://github.com/me-manu/crabmeyerpy}}.
The IC emissivity of the synchrotron emission involves four integrals for each frequency $\nu$ and radius $r$, see Eqs.~\eqref{eq:j-sync}, \eqref{eq:nseed}, and \eqref{eq:j-ic}.
In order to speed up the calculations, we make heavy use of numerical routines of \texttt{numpy}, \texttt{scipy}, and \texttt{numba}~\citep{Harris2020,Virtanen2020,Lam2015} and use multi-dimensional spline interpolations for $j^\mathrm{sync}_\nu$ and $n^\mathrm{sync}_\mathrm{seed}$.

The specific luminosity for component $t=($sync, dust, IC$)$ is found from the integration over radius, 
\begin{linenomath*}
\begin{equation}
    \mathcal{L}^{t}_\nu = \oint d\Omega \int dV j^t_\nu(\nu, r) =  (4\pi)^2\int\limits_{r_s}^{r_0}dr r^2 j_\nu^t(\nu, r)\,,
\end{equation}
\end{linenomath*}
where the two factors of $4\pi$ come from the volume and solid angle integrations, respectively. 
\begin{figure}
    \centering
	\includegraphics[width=0.99\linewidth]{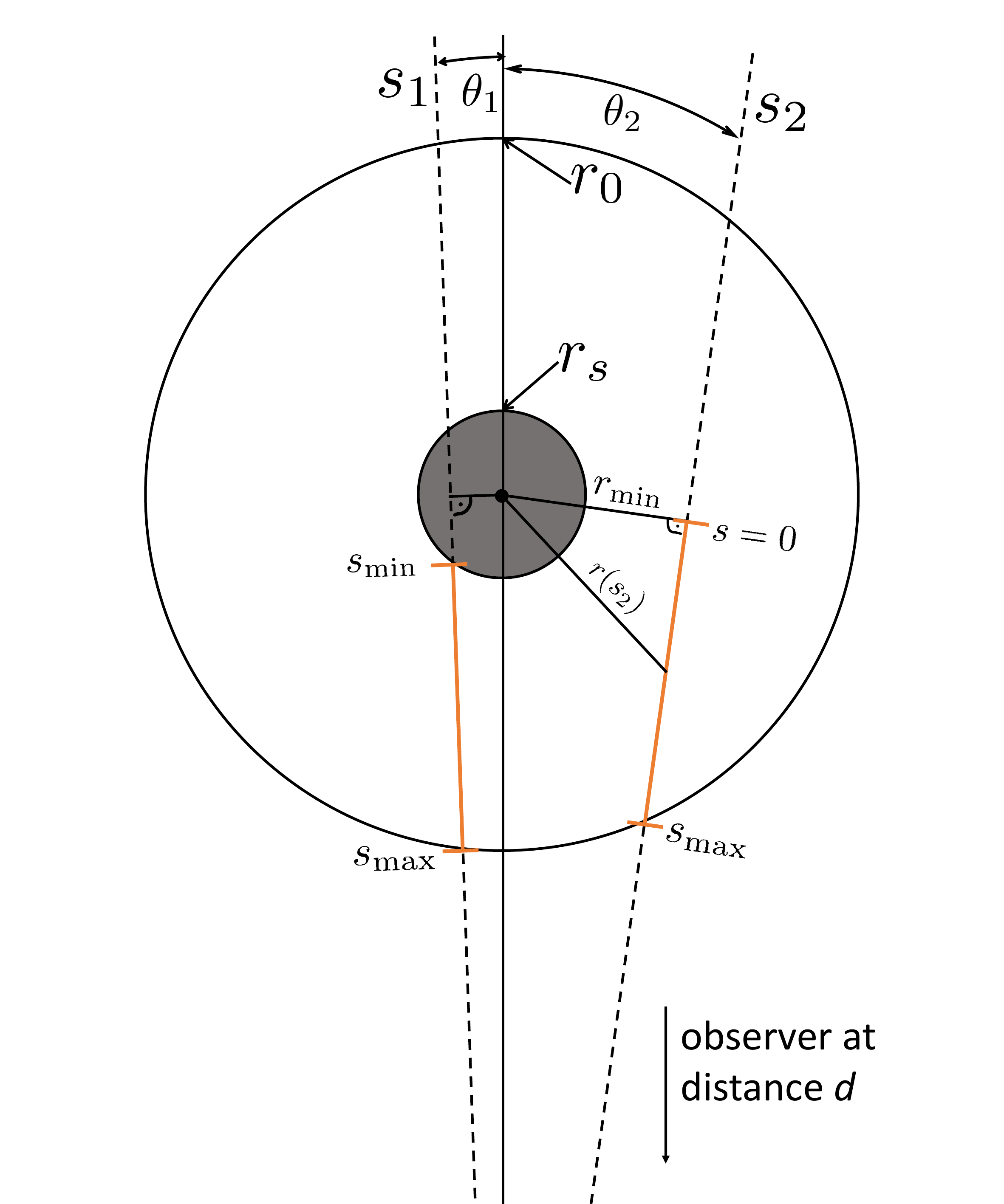}
	\caption{
        Illustration of the LOS integration.
        The pulsar in the centre is surrounded by a region with no emission, up to the shock radius $r_s$ (dark shaded region).
        We assume non-zero emission until the maximum radius of the nebula at $r_0$.
        Here we illustrate two exemplary lines of sight $s_1$ and $s_2$ at angular offsets $\theta_1$ and $\theta_2$ with respect to the centre (dashed lines).
        $s_1$ represents a line which intersects the shock whereas $s_2$ does not.
        We integrate the emissivity $j_\nu$ along the parts of the lines that are marked in solid orange.
        Because of the radial symmetry of our model, these parts contain exactly half of the emission expected from the full LOS.
	}
	\label{fig:los_scatch}
\end{figure}
For the analysis of \hess and \fermi data, we are also interested in the spatial profile of the emission for which we have to calculate the specific intensity $I_\nu$ as a function of the angular distance $\theta$ from the centre of the nebula. 
The intensity is given as an integral over the line of sight (LOS) $s$ ($s=0$ is the position on the LOS which is closest to the centre of the nebula) through the nebula in which $j_\nu^t$ is non-zero,
\begin{linenomath*}
\begin{equation}
    I^{t}_\nu(\theta) = 2\int\limits_{s_\mathrm{min}}^{s_\mathrm{max}}ds j_\nu^t(\nu, r(s))\,.
\end{equation}
\end{linenomath*}
This is illustrated in Fig.~\ref{fig:los_scatch}.
The distance $r(s)$ from each point on the LOS to the centre of the Crab Nebula calculates as  
\begin{linenomath*}
\begin{equation}
    r(s) = \sqrt{r_\mathrm{min}^2 + s^2}\,.
\end{equation}
\end{linenomath*}
In this expression, $r_\mathrm{min} = d\sin\theta$ is the smallest distance from the LOS to the centre of the Crab Nebula, which depends on the observing angle $\theta$ and the distance $d$ from the earth to the nebula.
Consequently, the emissivity along the LOS is symmetric with respect to the $s=0$ point which results in the factor two in front of the integral if we choose 
\begin{linenomath*}
\begin{equation}
    s_\mathrm{min} = \left\{
        \begin{array}{cl}
            \sqrt{r^2_s - r^2_\mathrm{min}} & r_\mathrm{min} \leq r_s \\
            0 & \, \textrm{otherwise}\\
        \end{array}.
    \right.
\end{equation}
\end{linenomath*}
The different cases correspond to a scenario where the LOS intersects with the shock radius and the integration starts at the point of the intersection, and another scenario where $\theta$ is large enough so that the LOS misses the shock radius.
The upper integration bound $s_\mathrm{max}$ is set to the intersection with $r_0$ where the emissivities are negligible small. 

The flux emitted from a point on the outer sphere of the nebula into a solid angle $\Omega$ is
\begin{linenomath*}
\begin{equation}
    F^{t}_\nu(\theta_0, \theta_1) = \int I_\nu \cos\theta d\Omega = 2\pi\int\limits_{\theta_0}^{\theta_1} I_\nu\cos\theta\sin\theta d\theta\,.
\end{equation}
\end{linenomath*}
The extension of our emission model is defined as the angle $\theta_{68}$ which contains 68\,\% of the flux, that is,
\begin{linenomath*}
\begin{equation}
    0.68 = \frac{F^t_\nu(0,\theta_{68})}{ F^t_\nu(0,\theta_\mathrm{max})}\,,
\end{equation}
\end{linenomath*}
where the maximum angle $\theta_{\max}$ is defined through $\tan\theta_{\max} = r_0 / d$. 

For a full description of the emission model, what remains to be defined are the electron spectrum $n_\mathrm{el}(\gamma,r)$ and the magnetic field profile $B(r)$. 
The tested models are described in the following subsections.

\subsection[Variable B-field model]{Variable $B$-field model}
\label{sec:var-B-model}
This is a phenomenological model developed in \citet{Meyer2010} and \citet{Dirson2023}, who expanded the previous works of \citet{deJager1992} and \citet{Hillas1998} to describe the Crab Nebula's SED and extension.
Two distinct electron populations are assumed: radio and wind electrons. 
The wind electrons are constantly injected by the pulsar wind and accelerated at the wind's termination shock, whereas the origin of the radio electrons is still unclear \citep{Atoyan1996}.
As the name suggests, the radio electrons are responsible for the synchrotron emission from radio frequencies to sub-millimetre / optical wavelengths whereas the wind electrons produce synchrotron emission at higher frequencies.
Both wind and radio electron densities are assumed to be zero for radii smaller than the shock radius.
The total electron spectrum is given by the sum of the two populations,
\begin{linenomath*}
\begin{equation}
    n_\mathrm{el}(\gamma, r) = n_\mathrm{radio}(\gamma, r) + n_\mathrm{wind}(\gamma, r)\,.
\end{equation}
\end{linenomath*}
The spectrum of the radio electrons is a simple power law with index $s_r$ between gamma factors $\gamma_{r,\mathrm{min}}$ and $\gamma_{r,\mathrm{max}}$ with a super-exponential cutoff,
\begin{linenomath*}
\begin{eqnarray}
    n_\mathrm{radio}(\gamma, r) & = & \frac{n_{r, 0}}{\rho_r^3} \gamma^{-s_r} \exp\left[-\left(\frac{\gamma}{\gamma_{r, \mathrm{max}}}\right)^{\beta_\mathrm{min}}\right] \nonumber \\
    & \times & \Theta(\gamma - \gamma_{r,\mathrm{min}}) \exp\left(-\frac{r^2}{2\rho_r^2}\right)\nonumber\\
    & \times & \Theta(r - r_s)\,,
\end{eqnarray}
\end{linenomath*}
where $\Theta(x)$ is the Heavyside step function.
The spatial dependence of the radio electrons is given by a Gaussian function of constant width $\rho_r$.
The wind electrons, on the other hand, are modelled with a double broken power law with super-exponential cut-offs at low and high energies.
As the radio wind electrons, the spatial extension of the wind electrons is assumed to follow a Gaussian whose width, however, also depends on energy.  
Putting everything together one finds
\begin{linenomath*}
\begin{eqnarray}
    n_\mathrm{wind}(\gamma, r) &=&  \frac{n_{w, 0}}{\rho_w(\gamma)^3}
    \exp\left[-\left(\frac{\gamma_{w, \mathrm{min}}}{\gamma}\right)^{\beta_\mathrm{min}}\right] \exp\left[-\left(\frac{\gamma}{\gamma_{w, \mathrm{max}}}\right)^{\beta_\mathrm{max}}\right]
    \nonumber\\
    &\times & \left[\left(\frac{\gamma}{\gamma_{w,1}}\right)^{-s_{w,1}}
    \left(\frac{\gamma_{w,1}}{\gamma_{w,2}}\right)^{-s_{w,2}}
    (1-\Theta(\gamma - \gamma_{w,1}))\right. \nonumber\\
    & + &  \left(\frac{\gamma}{\gamma_{w,2}}\right)^{-s_{w,2}}
    \mathcal{B}(\gamma, \gamma_{w,1}, \gamma_{w,2}) \nonumber \\
    & + &  \left.\left(\frac{\gamma}{\gamma_{w,2}}\right)^{-s_{w,3}}
    \Theta(\gamma - \gamma_{w,2})\right]\nonumber\\
    &\times&\exp\left(-\frac{r^2}{2\rho_w(\gamma)^2}\right)\nonumber\\
    & \times & \Theta(r - r_s)\,,
\end{eqnarray}
\end{linenomath*}
where $\mathcal{B}(x, a, b)$ is the boxcar function, $\mathcal{B}(x, a, b)=\Theta(x - a) - \Theta(x - b)$.
The energy dependence of the spatial extension is modelled through a simple power law
\begin{linenomath*}
\begin{equation}
\label{eq:rho_w}
    \rho_{w} (\gamma) = \rho_{w,0}\left[\left(\frac{\gamma}{9\cdot10^5}\right)^2\right]^{-\alpha_w}\,.
\end{equation}
\end{linenomath*}
For the free parameters $\gamma_\mathrm{r,max}$, $\gamma_\mathrm{w,min}$, and $\gamma_\mathrm{w,max}$ we do not use abrupt cutoffs as those would lead to discontinuities on the numerical integration grid.
Instead we use super-exponential cutoffs which are not motivated physically but help the smoothness of the likelihood landscape.
For simplicity, the super-exponential indices are fixed to $\beta_\mathrm{min} = 2.8$~\citep[this is the best-fit value obtained by][]{Meyer2010} and  $\beta_\mathrm{max} = 2$.

For the magnetic field profile, we follow \citet{Dirson2023} and choose 
\begin{linenomath*}
\begin{equation}
    B(r) = B_0\left(\frac{r}{r_s}\right)^{-\alpha}\,,
\end{equation}
\end{linenomath*}
with $\alpha \geqslant 0$. 
The case of $\alpha=0$ corresponds to the constant $B$-field model studied, for example, in \citet{Meyer2010}.

\subsection{MHD flow model}
This model follows the solution to the steady-state MHD equations under the assumption of spherical symmetry and describes the flow of the electron plasma from the shock to the nebula's boundary. 
The MHD flow and the magnetic field are determined by the magnetisation $\sigma$ at the shock, which is given by the ratio of the electromagnetic and particle energy flux, 
\begin{linenomath*}
\begin{equation}
    \sigma = \frac{B_s^2}{4\pi n u_s \gamma m_e c^2}\,,
\end{equation}
\end{linenomath*}
where $B_s$ is the magnetic field, $n$ the particle density, and $u_s=\sqrt{\gamma^2 - 1}$ the relativistic four speed (all quantities are at the shock radius). 

The (injected) electron spectrum is another free parameter in the model. 
As in the previous model, we use two distinct electron populations, with the radio electrons modelled in the same way as in Section~\ref{sec:var-B-model}. 
On the other hand, the spatial and spectral distribution for the wind electrons are the result of solving the MHD equations for some electron population injected at the termination shock. 
While \citet[][]{Kennel1984} considered a power law type injection, this was generalised by \citet{Atoyan1996} to a more complex shape, which we also assume here with an additional break in the spectrum:
\begin{linenomath*}
\begin{eqnarray}
    n_{s}(\gamma) &=& n_{0, w} \exp\left[-\left(\frac{\gamma_{w, \mathrm{min}}}{\gamma}\right)^{\beta_\mathrm{min}}\right] 
    \exp\left[-\left(\frac{\gamma}{\gamma_{w, \mathrm{max}}}\right)^{\beta_\mathrm{max}}\right]  \nonumber\\
    & \times & \left[\left(1+\frac{\gamma}{\gamma_{w,\mathrm{min}}}\right)^{s_{w, 1}} (1-\Theta(\gamma - \gamma_{w,2}))\right. \nonumber \\
    & + & \left.\frac{\left(1 + \gamma/\gamma_{w,\mathrm{min}}\right)^{s_{w,3}}}{\left(1+\gamma_{w,2}/\gamma_{w,\mathrm{min}}\right)^{s_{w,3}-s_{w,1}}} \Theta(\gamma - \gamma_{w,2}))\right]\,.
\end{eqnarray}
\end{linenomath*}
Similar to the wind electron spectrum for the static models, $\beta_\mathrm{min}$ and $\beta_\mathrm{max}$ are fixed to 2.8 and 2.0, respectively.
From the MHD flow solution, the energy of electrons at some distance $z = r/r_s$ from the shock can be related to the electron Lorentz factor at injection, $\gamma'$,
\begin{linenomath*}
\begin{equation}
    \gamma(z) = \frac{\gamma'}{(vz^2)^{1/3} + \gamma'/\gamma_\mathrm{max}}\,,
\end{equation}
\end{linenomath*}
where $v(z) = u(z) / u_s$ is the four-velocity of the electrons \citep[see Eqs. (A7a) - (A7d) in][]{Kennel1984a} which depends on the magnetisation $\sigma$, and $\gamma_\mathrm{max}$ is the maximum $\gamma$ factor at distance $z$, see, for example,  Eq.~(6) in \citet{Atoyan1996} (which also depends on $vz^2$). 
The wind electron spectrum is then found to be 
\begin{linenomath*}
\begin{equation}
    n_\mathrm{wind}(\gamma, z) = (vz^2)^{-4/3}\left(\frac{\gamma'}{\gamma(z)}\right)^2 n_s(\gamma')\,.
\end{equation}
\end{linenomath*}

For small values of the magnetisation, the magnetic field in the MHD solution at the shock is given by 
\begin{linenomath*}
\begin{equation}
    B_s = \sqrt{\frac{L_\mathrm{spin-down}}{c r_s^2}\frac{\sigma}{1 + \sigma}}\,,
\end{equation}
\end{linenomath*}
where $L_\mathrm{spin-down}$ is the spin-down luminosity of the pulsar.
Downstream of the shock, the magnetic field in the nebula evolves as
\begin{linenomath*}
\begin{equation}
    B(z) = B_s  \frac{3(1 - 4\sigma)z}{vz^2}\,.
\end{equation}
\end{linenomath*}

\newpage

\section{Best-fit parameters of physical models}
\label{sec:appendix_fit_parameters}
We present the best-fit parameters for all considered models in Table~\ref{table:parameters}. 
The best-fit values were found using the `simplex' method of the \textsc{sherpa} fitting backend of \gp. 
The Nelder-Mead algorithm converged into slightly lower minima than the `migrad' method from the \textsc{minuit} backend.
As we found the estimation of uncertainties on the fit parameters to not work reliably -- presumably due to the complexity of the model and the large number of parameters -- we specify only the best-fit values here.

\begin{table*}
    \centering
    \caption{
        Best-fit parameter values of the physical models.
        Parameters that are not free in the fit are indicated by `(fixed)' after their value.
    }
    \label{table:parameters}
    \begin{tabular}{l c c c}
        \hline\hline
        Parameter & \vbm  &  \cbm & Kennel \& Coroniti\\
        \hline
$\ln(n_{r,0})$ & $117.170$ & $117.69$ & $118.766$\\
$\ln(\gamma_{r,\mathrm{min}})$ & $3.09$ (fixed) & $3.09$ (fixed) & $3.09$ (fixed)\\
$\ln(\gamma_{r,\mathrm{max}})$ & $11.599$ & $12.35$ & $12.625$\\
$s_r$ & $-1.5439$ & $-1.649$ & $-1.7419$\\
$\rho_r\,['']$ & $88.3$ & $80.40$ & $88.64$\\ \hline
$\ln(n_{w,0})$ & $76.822$ & $76.8315$ & $-27.625$\\
$\ln(\gamma_{w,\mathrm{min}})$ & $12.841$ & $12.69$ & $12.8366$\\
$\ln(\gamma_{w,1})$ & $15.26$ & $14.24$ &  --- \\
$\ln(\gamma_{w,2})$ & $19.197$ & $19.35379$ & $17.96$\\
$\ln(\gamma_{w,\mathrm{max}})$ & $22.115$ & $22.371$ & $22.251$\\
$\beta_\mathrm{min}$ & $2.8$ (fixed) & $2.8$ (fixed) & $2.8$ (fixed)\\
$\beta_\mathrm{max}$ & $2$ (fixed) & $2$ (fixed)  & $2$ (fixed) \\
$s_{w,1}$ & $-3.117$ & $-2.75$ &  --- \\
$s_{w,2}$ & $-3.3928$ & $-3.1764$ & $-2.8695$\\
$s_{w,3}$ & $-3.782$ & $-3.5118$ & $-2.316$\\
$\rho_{w,0}\, ['']$ & $98.14$ & $78.94$ &  --- \\
$\alpha_w$ & $0.12544$ & $0.11973$ &  --- \\ \hline
$B_0\, [\si{\micro\gauss}]$ & $256.4$ & $126.39$ &  --- \\
$r_\mathrm{s}\, ['']$ & $13.4$ (fixed) & $13.4$ (fixed) & $13.4$ (fixed)\\
$\alpha$ & $-0.4691$ &  ---  &  --- \\ \hline
$\sigma$ &  ---  &  ---  & $0.021396$\\
$\ln(L_\mathrm{spin-down} [\mathrm{erg/s}])$ &  ---  &  ---  & $88.716$\\ \hline
$r_\mathrm{dust, in}$ [pc] & $0.55$ (fixed) & $0.55$ (fixed) & $0.55$ (fixed)\\
$r_\mathrm{dust, out}$ [pc] & $1.53$ (fixed) & $1.53$ (fixed) & $1.53$ (fixed)\\
$\log_{10}(M_1/M_\odot)$ & $-4.4$ (fixed) & $-4.4$ (fixed) & $-4.4$ (fixed)\\
$\log_{10}(M_2/M_\odot)$ & $-1.2$ (fixed) & $-1.2$ (fixed) & $-1.2$ (fixed)\\
$T_1$ [K] & $149$ (fixed) & $149$ (fixed) & $149$ (fixed)\\
$T_2$ [K] & $39$ (fixed) & $39$ (fixed) & $39$ (fixed)\\
        \hline
    \end{tabular}
    \tablefoot{For an explanation of the parameters, see Appendix~\ref{appx-emission_model}.}
\end{table*}

\end{appendix}

\end{document}